# Coupling-Aware Aggregation of Multi-Zone HVAC Loads under Uncertainty: A Two-level Framework

Jingguan Liu, *Graduate Student Member, IEEE,* Han Jiang, *Member, IEEE*, Xiaomeng Ai, *Member, IEEE,* Shengshi Wang*, Member, IEEE,* Xizhen Xue, *Member, IEEE*, Shichang Cui, *Member, IEEE,* Jinming Hou, Jiakun Fang, *Senior Member, IEEE*, and Jinyu Wen*, Senior Member, IEEE*

***Abstract*—Aggregating building heating, ventilation, and air-conditioning (HVAC) loads unlocks substantial demand-side flexibility for power systems. Yet multi-zone coupling creates intricate interdependencies and uncertainty propagation, complicating the quantification of aggregate flexibility. To address this issue, this paper proposes a coupling-aware two-level aggregation framework. At the building level, tailored Gaussian elimination and coordinate transformation techniques are employed to recast the high-dimensional thermal dynamics as an equivalent lower-dimensional analytical expression. This expression streamlines the subsequent aggregator-level stage by (i) clarifying the propagation of zone-level uncertainties to the building-level interface, (ii) decoupling intra-building multi-zone coupling from inter-building aggregation, and (iii) providing full-dimensional building-level flexibility sets that enable tractable reformulation. At the aggregator level, existing geometric aggregation approaches are generalized by a newly developed matrix-transformation technique. This technique effectively constructs inner approximations between polytopes of different dimensions, producing closed-form images of high-dimensional multi-zone HVAC flexibility in power subspace. The resulting inner approximation is then recast as a customized separatable linear program that efficiently determines the optimal aggregate parameters. Case studies validate the effectiveness of our framework, highlighting its accuracy, reliability, and scalability.**

## NOMENCLATURE

Main symbols and notations used in this paper are given as follows, with additional ones defined as needed. Scalars are written in italic, matrices and vectors in boldface, and sets in blackboard bold. The all-ones, all-zeros, and identity matrices are denoted by $\mathbf{1}$, $\mathbf{0}$, and $\mathbf{I}$, respectively. The vertical and horizontal concatenations of matrices $\mathbf{A}$ and $\mathbf{B}$ are denoted by $[\mathbf{A};\mathbf{B}]$ and $[\mathbf{A},\mathbf{B}]$, respectively. The inverse, transpose, trace, and rank of a matrix $\mathbf{A}$ are expressed as $(\mathbf{A})^{-1}$, $(\mathbf{A})^{\mathrm{T}}$, $\mathrm{tr}(\mathbf{A})$, and $\mathrm{rank}(\mathbf{A})$, respectively. The dimension of a vector $\mathbf{b}$ is denoted by $\dim(\mathbf{b})$.

*A. Abbreviation*

| | |
|---|---|
| DER | Distributed energy resource. |
| DRCC | Distributionally robust chance constraint. |
| HVAC | Heating, ventilation, and air-conditioning. |
| H-representation | Half-space representation. |
| UPR | Unused potential ratio. |
| VB | Virtual battery. |

*B. Indices:*

| | |
|---|---|
| $t$ | Dispatch period index. |
| $i/j$ | Zone index. |
| $n$ | Building index. |

*C. Sets:*

| | |
|---|---|
| $\mathbb{U}^{\mathrm{hvac}}$ | Zone-level HVAC flexibility set. |
| $\mathbb{U}^{\mathrm{bld}}$ | Building-level HVAC flexibility set. |
| $\mathbb{P}^{\mathrm{bld}}$ | Projection of $\mathbb{U}^{\mathrm{bld}}$ onto the power subspace. |
| $\mathbb{P}^{\mathrm{agg}}$ | Exact aggregator-level aggregate set. |
| $\mathbb{P}^{\mathrm{app}}$ | Approximate aggregator-level aggregate set. |
| $\mathbb{P}^{\mathrm{base}}$ | Base set. |
| $\mathbb{P}^{\mathrm{aff}}$ | Affine-transformed base set. |

*D. Parameters:*

| | |
|---|---|
| $N^{\mathrm{T}}$ | Dispatch period number |
| $N^{\mathrm{I}}$ | Zone number |
| $N^{\mathrm{B}}$ | Building number. |
| $\Delta t$ | Period length. |
| $R^{\mathrm{in}}$ | Thermal resistance between indoor temperature nodes including zone nodes and wall nodes. |
| $R^{\mathrm{out}}$ | Thermal resistance between indoor temperature node and outdoor temperature node. |
| $C^{\mathrm{in}}$ | Thermal capacity of indoor temperature node. |
| $\pi^{\mathrm{out}}$ | Outdoor connection identifier, which is equal to 1 for indoor temperature nodes connected to the outdoor temperature node, and 0 otherwise. |

This work was supported by the National Natural Science Foundation of China under Grant 52177088 and Grant 52207108 *(Corresponding author: Xiaomeng Ai).*

J. Liu, X. Ai, S. Cui, J. Fang, and J. Wen are with the State Key Laboratory of Advanced Electromagnetic Technology, Huazhong University of Science and Technology, Wuhan 430074, China (e-mail: spencerplusmail@foxmail.com; xiaomengai@hust.edu.cn; shichang_cui@hust.edu.cn; jfa@hust.edu.cn; jinyu.wen@hust.edu.cn).

S. Wang is with the Engineering Cluster, Singapore Institute of Technology, Singapore 828608 (e-mail: sensewang1997@gmail.com).

X. Xue is with the School of Electrical and Electronic Engineering, Nanyang Technological University, 639798, Singapore (email: xizhen.xue@ntu.edu.sg).

H. Jiang and J. Hou are with the Global Energy Interconnection Group Co., Ltd., Beijing 100031, China (email: han-jiang@geidco.org; jinming-hou@geidco.org).

$T^{\text{out}}$ Outdoor temperature.
$\pi^{\text{int}}$ Internal heat identifier, which is equal to 1 for indoor zone nodes, and 0 otherwise.
$Q^{\text{int}}$ Internal heat gain.
$\pi^{\text{hvac}}$ HVAC identifier, which is equal to 1 for indoor zone nodes with HVAC, and 0 otherwise.
$\eta^{\text{hvac}}$ Coefficient of performance for HVAC unit.
$T^{\text{set}}$ Indoor zone temperature set-point
$\beta^{\text{set}}$ Indoor zone temperature tolerance.
$P^{\text{hvac,max}}$ Maximum power of HVAC unit.
$P^{\text{hvac,min}}$ Minimum power of HVAC unit.

### *E. Decision Variables:*

$T^{\text{in}}$ Indoor zone temperature and wall temperature.
$P^{\text{hvac}}$ Cooling power of HVAC unit.

## I. Introduction

COORDINATED control of demand-side distributed energy resources (DERs) can substantially enhance power system flexibility [1], [2]. Among these resources, building heating, ventilation, and air conditioning (HVAC) loads are especially attractive because of their fast response and significant potential to shift consumption via pre-cooling or pre-heating [3]. However, scheduling each HVAC load individually becomes computationally prohibitive at scale [4]. *Aggregators* address this challenge by contracting with numerous HVAC loads, aggregating their flexibility into a single flexibility set for grid operators, and then dispatching the approved power trajectory to individual units [5]. Therefore, for aggregators, accurate and reliable quantification of aggregate flexibility is essential.

### *A. Related Work*

Mathematically, the flexibility of an individual HVAC load can be represented as a subset of the power space, and the aggregate flexibility corresponds to the Minkowski sum of these subsets [6]. Since exact computation of this sum is generally NP-hard [7], most research relies on closed-form approximations. Inner approximations are particularly attractive because they retain only feasible operating points, ensuring safe operation [8]. Inner approximation techniques fall into two main categories:

*1) Device-wise Geometric Approaches:* These approaches first approximate each DER by an inner set, then derive the Minkowski sum of these approximations in closed form. Common approximations include zonotopes [9], [10], homothetic polytopes [1], [6], [7], and affine polytopes [3], [8], [11], [12]. Among these options, affine polytopes demonstrate excellent geometric adaptability for capturing the complex thermal dynamics in HVAC systems [3]. These approaches perform well with uncoupled DERs. Here, each flexibility set projects directly onto the power subspace, with its inner polytope remaining within that subspace. However, multi-zone HVAC loads violate this assumption. The multi-zone thermal interactions create interdependent, high-dimensional flexibility sets whose projection onto the power subspace becomes NP-hard. Popular projection algorithms such as Fourier–Motzkin elimination rapidly encounter the curse of dimensionality [13]. Hence, this multi-zone coupling creates dimensional mismatches between HVAC flexibility sets and their inner approximations in the power subspace, rendering existing device-wise geometric approaches inapplicable.

*2) One-step Projection Approaches:* These approaches reformulate the Minkowski sum as a projection of all DER sets in one step. Since exact projection is NP-hard, these approaches employ simplified templates such as hyper-boxes [14], [15], ellipsoids [16], [17], and virtual batteries (VBs) [18], [19]. By bypassing the explicit Minkowski-sum computation, these approaches are able to accommodate the multi-zone coupling. However, this advantage comes at the cost of two limitations. First, prespecified simple templates cannot adapt to complex thermal coupling patterns, comprising aggregation accuracy. Second, these approaches suffer from poor scalability. Since all HVAC flexibility sets must be processed simultaneously, computational costs increase rapidly with the number of HVAC loads, preventing large-scale deployment [20].

In summary, existing approaches face a fundamental trade-off: they must either impose restrictive dimensional consistency for efficient aggregation or accommodate coupling at the cost of scalability. Thus, developing a computationally efficient aggregation framework that handles multi-zone coupling remains an open challenge.

In addition to dimensional mismatch, the multi-zone coupling propagates uncertainties across zones, undermining aggregation reliability [21], [22]. Recent studies have observed that fluctuations in outdoor temperatures and indoor heat gains can markedly alter HVAC operating envelopes [23], [24]. Ignoring these effects yields unreliable flexibility estimates and can even render disaggregation infeasible when constraints are violated [25]. To hedge against forecast errors, more advanced methods have been proposed for constructing more reliable aggregate sets. For example, the robust optimization method is proposed in [26] to safeguard against worst-case scenarios but tend to be overly conservative. In [7], [27], [28], the distributionally robust chance constraints (DRCCs) are adopted to offer a better balance between conservatism and reliability. Nevertheless, existing methods typically treat uncertainties as independent perturbations on each DER, failing to capture uncertainty propagation through thermal interactions among coupled zones to the aggregate power interface. Quantifying these propagation pathways could greatly enhance the reliability of aggregated flexibility and help aggregators manage risk, yet this aspect remains largely unexplored.

### *B. Main Contributions and Paper Organization*

In light of the identified research gaps, this paper pioneers a two-level framework for coupling-aware aggregation of large-scale multi-zone HVAC loads under uncertainty. At the building level, zone-level thermal couplings are mapped to a building-level power interface and the uncertainty propagation is explicitly quantified. At the aggregator level, these building-level expressions are integrated to yield an accurate and reliable aggregate set for a large population of buildings.

TABLE I
COMPARISONS OF INNER APPROACHES FOR AGGREGATING MULTI-ZONE HVAC LOADS

| Approach | Multi-zone-coupled aggregation | Computation complexity for large-scale implementation | Aggregation accuracy | Uncertainty propagation |
|---|---|---|---|---|
| Zonotope [9], [10] | Inapplicable | Low | Medium | Ignore |
| Homothet polytope [1], [6], [7] | Inapplicable | Low | Medium | Ignore |
| Affine polytope [3], [8], [11] | Inapplicable | Low | High | Ignore |
| Ellipsoid [16], [17] | Applicable | High | Medium | Ignore |
| Hyper-box [14], [15] | Applicable | Low | Low | Ignore |
| Virtual battery [18], [19] | Applicable | High | Medium | Ignore |
| **Ours** | **Applicable** | **Low** | **High** | **Handled** |

Our main contributions are summarized below, with comparisons to existing studies provided in TABLE I:

*1) New Expression*: We develop a novel concise analytical expression for multi-zone thermal coupling within buildings. Through tailored Gaussian elimination and coordinate transformation, complex high-dimensional thermal dynamics are equivalently reduced to a lower-dimensional expression at the building-level power interface. This expression streamlines the subsequent aggregation by: (i) clarifying uncertainty propagation from zones to the building interface, (ii) decoupling intra-building dynamics from inter-building aggregation, and (iii) providing full-dimensional building-level flexibility sets that enable tractable reformulation.

*2) Generalized Approach:* To the best of our knowledge, this is the first time to extend existing polytope-based aggregation approaches [1], [3], [6], [7], [8], [11] to handle multi-zone coupling. Leveraging the derived building-level expression and a new matrix-transformation technique, we construct inner approximations between polytopes of different dimensions. This produces closed-form representations of high-dimensional HVAC flexibility in power subspace. We further reformulate the inner approximation as a customized tractable, separable linear program which determines the optimal aggregate parameters and ensures accurate, reliable, and efficient aggregation of large-scale HVAC loads under uncertainty.

Note that while this paper has focused on coupling-aware aggregation of multi-zone HVAC loads, the proposed framework remains applicable, with minor modifications, to other DER resources whose coupled flexibility sets can similarly be expressed or approximated by convex polytopes in half-space representations—such as data center loads coupled through computational task allocation [29].

The remainder of the paper is organized as follows. Section II outlines the problem formulation. Section III describes the proposed two-level aggregation framework. Section IV presents case studies with conclusion in Section V.

## II. PROBLEM FORMULATION

This section introduces the mathematical model of building multi-zone HVAC loads and formulates the problem of finding inner approximations of the aggregate flexibility set.

### A. Multi-Zone-Coupled Modeling

Fig. 1 illustrates a typical schematic of multi-zone HVAC loads in a building. The indoor environments exchange heat with the ambient environment and adjacent zones. The thermal dynamics of the temperature nodes $T_{n,i,t}^{\text{in}}$ of zone $i$, which interact with the adjacent nodes $T_{n,j,t}^{in}$ of zone $j$, can be described using a finite difference formulation [21], [30], [31]:

$$C_{n,i}^{\text{in}} \frac{T_{n,i,t}^{\text{in}} - T_{n,i,t-1}^{\text{in}}}{\Delta t} = \sum_{j,j \neq i} \frac{T_{n,j,t}^{\text{in}} - T_{n,i,t}^{\text{in}}}{R_{n,ij}^{in}} + \pi_{n,i}^{\text{out}} \frac{T_{n,i,t}^{\text{out}} - T_{n,i,t}^{\text{in}}}{R_{n,i}^{out}} + \pi_{n,i}^{\text{int}} Q_{n,i,t}^{\text{int}} + \pi_{n,i}^{\text{hvac}} \eta_{n,i}^{\text{hvac}} P_{n,i,t}^{\text{hvac}} \tag{1.a}$$

To ensure the users' thermal comfort, both the indoor zone temperature limit and the HVAC power limit should be considered, as shown in (1.b)-(1.c).

$$T_{n,i}^{\text{set}} - \beta_{n,i}^{\text{set}} \leq T_{n,i,t}^{\text{in}} \leq T_{n,i}^{\text{set}} + \beta_{n,i}^{\text{set}} \tag{1.b}$$

$$P_{n,i}^{\text{hvac,min}} \leq P_{n,i,t}^{\text{hvac}} \leq P_{n,i}^{\text{hvac,max}} \tag{1.c}$$

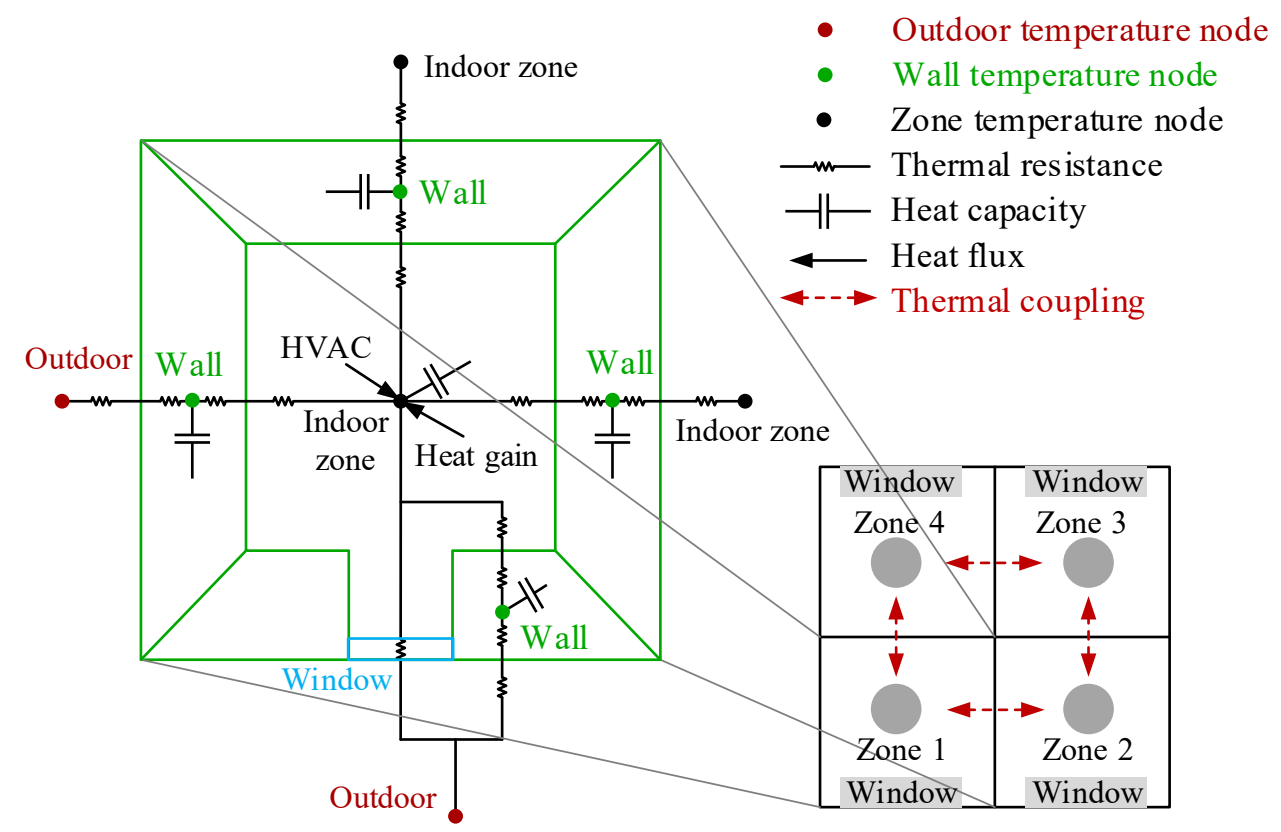

Fig. 1. Illustration of multi-zone-coupled thermal dynamics.

Combining (1.a)-(1.c), the feasible region for multi-zone HVAC loads in building $n$ is expressed as the half-space representation (H-representation) of a polytope $\mathbb{U}_n^{\text{hvac}}$ in the space defined by node temperature vector $\mathbf{T}_n^{\text{in}} \in \mathbb{R}^{N_n^{\text{I}} N^{\text{T}}}$ and HVAC power vector $\mathbf{P}_n^{\text{hvac}} \in \mathbb{R}^{N_n^{\text{I}} N^{\text{T}}}$. Here, the bold variable vectors denote the column vector formed by stacking the corresponding elements column-wise, e.g., $\mathbf{T}_n^{\text{in}} := \left[T_{n,1,1}^{\text{in}}; T_{n,2,1}^{\text{in}}; \ldots; T_{n,N_n^{\text{I}},1}^{\text{in}}; T_{n,1,2}^{\text{in}}; T_{n,2,2}^{\text{in}}; \ldots; T_{n,N_n^{\text{I}},N^{\text{T}}}^{\text{in}}\right]$.

The thermal dynamics in (1.a) are determined by a series of parameters, among which outdoor temperature $T_{n,i,t}^{\text{out}}$ and internal heat gain $Q_{n,i,t}^{\text{int}}$ are uncertain parameters obtained through prediction. These parameters vary during real-time dispatch and greatly affect $\mathbb{U}_n^{\text{hvac}}$, thereby being considered as the uncertain parameters denoted by $\boldsymbol{\xi}_n := [\mathbf{T}_n^{\text{out}}; \mathbf{Q}_n^{\text{int}}]$. We are now ready to express $\mathbb{U}_n^{\text{hvac}}(\boldsymbol{\xi}_n)$ in a compact form as shown in (2). Here, $\mathbf{A}^{(\cdot)}/\mathbf{b}^{(\cdot)}$ is the coefficient matrix/vector parameter derived from (1.a)-(1.c).

$$\mathbb{U}_n^{\text{hvac}}(\xi_n) := \left\{ \left[ \mathbf{T}_n^{\text{in}}; \mathbf{P}_n^{\text{hvac}} \right] \middle| \begin{array}{l} \mathbf{A}_n^{\text{eq1}}\mathbf{T}_n^{\text{in}} + \mathbf{A}_n^{\text{eq2}}\mathbf{P}_n^{\text{hvac}} + \mathbf{A}_n^{\text{eq3}}\xi_n = \mathbf{b}_n^{\text{eq}} \\ \mathbf{A}_n^{\text{ieq1}}\mathbf{T}_n^{\text{in}} \le \mathbf{b}_n^{\text{ieq1}}, \mathbf{A}_n^{\text{ieq2}}\mathbf{P}_n^{\text{hvac}} \le \mathbf{b}_n^{\text{ieq2}} \end{array} \right\} \quad (2)$$

### B. Aggregate Flexibility Set

For an aggregator controlling $N^{\text{B}}$ buildings, the aggregate flexibility set $\mathbb{P}^{\text{agg}}(\boldsymbol{\xi}_n) \subseteq \mathbb{R}^{N^{\text{T}}}$ can be expressed as the Minkowski sum of the HVAC power in each zone as below:

$$\mathbb{P}^{\text{agg}}(\xi_n) := \left\{ \mathbf{P}^{\text{agg}} \middle| \underbrace{P_t^{\text{agg}} = \sum_n \sum_i P_{n,i,t}^{\text{hvac}}}_{\text{Minkowski sum}}, \underbrace{P_{n,i,t}^{\text{hvac}} \in \mathbb{U}_n^{\text{hvac}}(\xi_n)}_{\text{multi-zone coupling}} \right\} \quad (3)$$

In general, computing the Minkowski sum of two arbitrary H-representation polytopes is NP-hard [32], [33]. Multi-zone coupling further complicates this challenge by introducing interdependencies among $P_{n,i,t}^{\text{hvac}}$. Additionally, such coupling propagates uncertainties across zones, making it difficult to quantify their cumulative effect on aggregation.

In light of these challenges, we primarily aim to achieve an accurate and reliable aggregation via inner approximation of $\mathbb{P}^{\text{app}}$. Specifically, our objective is to identify an approximate polytope $\mathbb{P}^{\text{app}}$ such that $\mathbb{P}^{\text{app}} \subseteq \mathbb{P}^{\text{agg}}(\boldsymbol{\xi}_n)$ holds.

## III. Two-Level Aggregation Framework

This section introduces the proposed two-level aggregation framework to find the inner approximation of the aggregate set.

### A. Framework Overview

While computing the Minkowski sum of multi-zone HVAC flexibility sets is challenging, fortunately, coupling occurs only within individual buildings—zones from different buildings remain independent. Leveraging this key insight (Fig. 2), our framework decomposes the aggregation problem into building and aggregator levels to achieve computational efficiency.

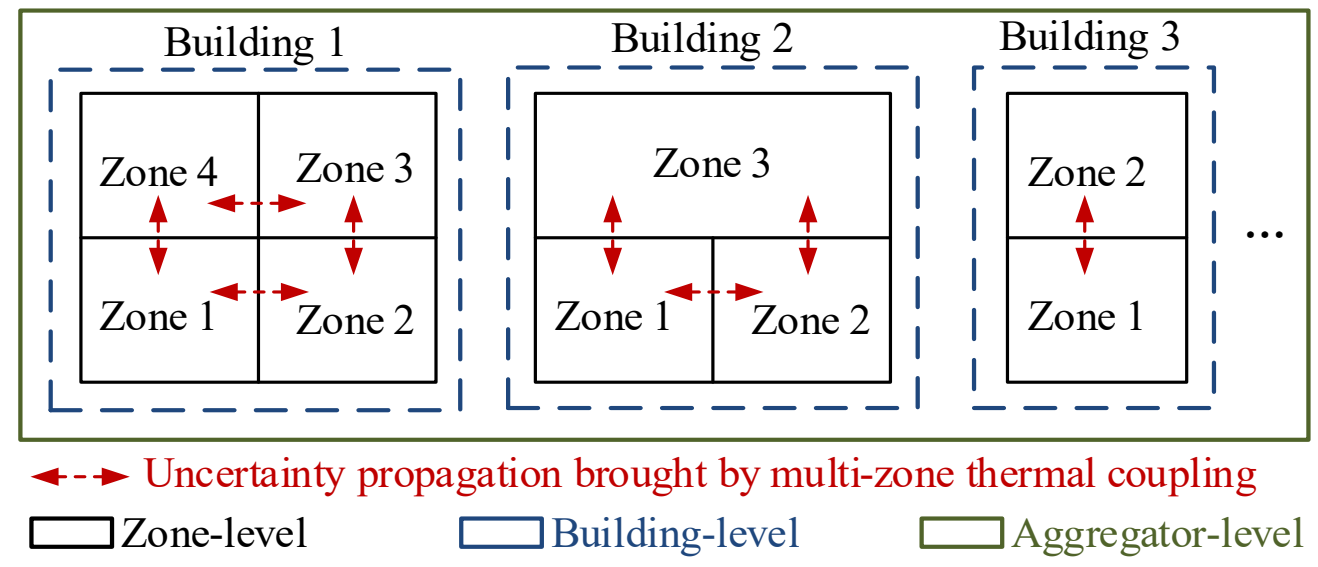

Fig. 2. Relationship of different levels.

The detailed components of our framework are as follows:

1) At the building level, we develop a concise analytical expression at the building-level power interface. The expression captures intra-building multi-zone thermal coupling and associated uncertainty propagation, streamlining the subsequent aggregator-level stage (Section III.B).

2) At the aggregator level, we leverage the building-level expression and matrix-transformation technique to construct inner approximation of high-dimensional multi-zone HVAC flexibility in power subspace. The approximation enables closed-form calculation of Minkowski sum (Section III.C).

3) We formulate a linear program to efficiently determines the optimal aggregate parameters for inner approximations, ensuring accurate, reliable, and efficient aggregation of large-scale HVAC loads under uncertainty (Section III.D).

4) Leveraging the aggregate parameters, we design a fast and feasible disaggregation strategy for efficient power dispatch of large-scale building HVAC loads (Section III.E).

To show the logical relationships among the individual components, Fig. 3 presents the overall sequential workflow of our framework for the aggregator's day-ahead implementation.

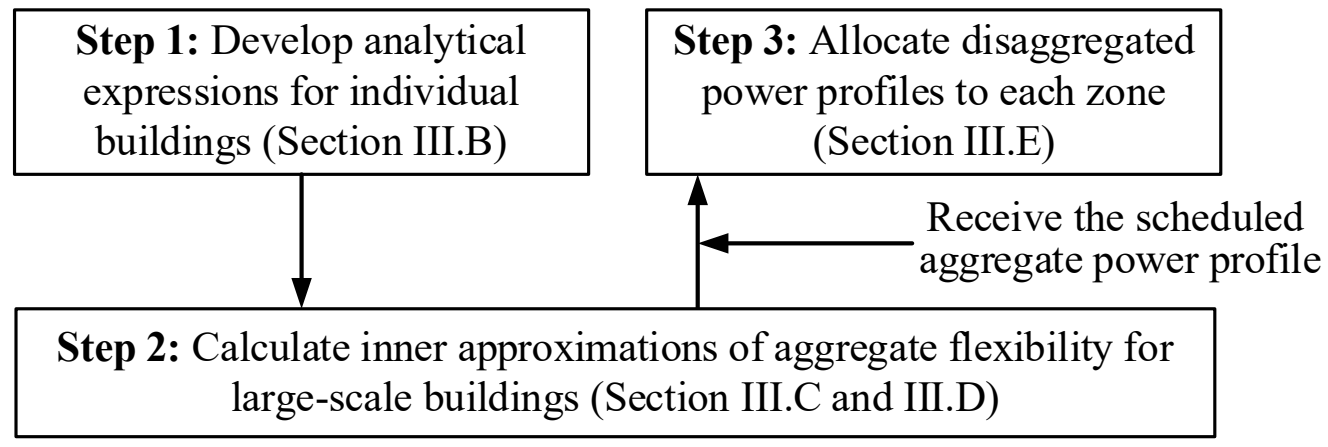

Fig. 3. Overall sequential workflow of our framework for aggregators.

### B. Building-Level Analytical Expression

At the building level, we derive a compact analytical expression $\mathbb{U}_n^{\text{bld,low}}(\boldsymbol{\xi}_n)$ that characterizes the multi-zone thermal interactions within a building. The expression is obtained through three key steps that project the high-dimensional thermal dynamics $\mathbb{U}_n^{\text{bld,hig}}(\boldsymbol{\xi}_n)$ onto an equivalent reduced subspace (Fig. 4).

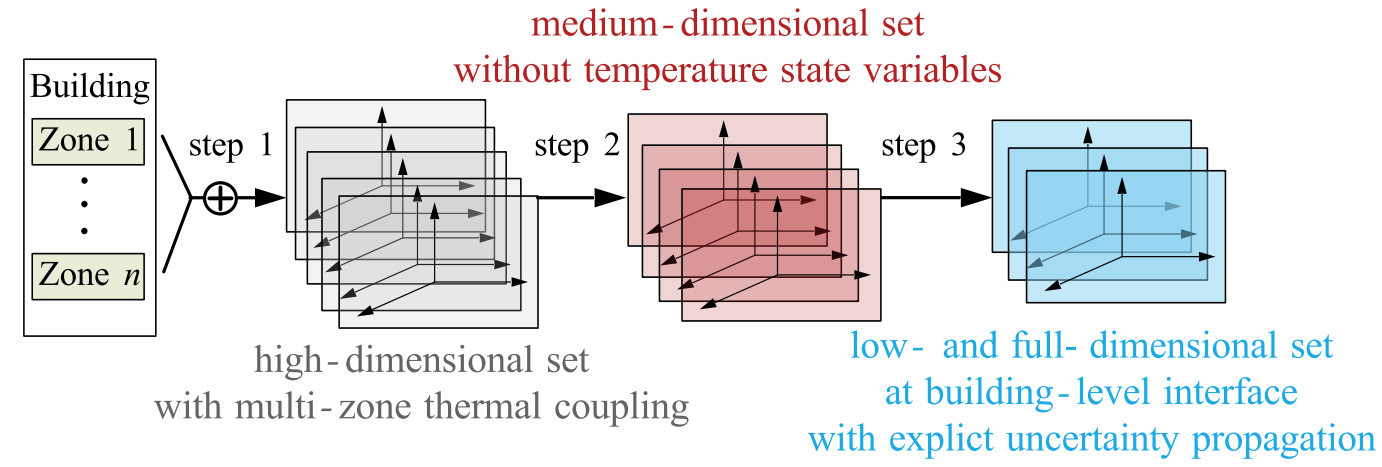

Fig. 4. Illustration of building-level projection.

**Step 1** (Lifting to a high-dimensional space): We construct the high-dimensional polytopes $\mathbb{U}_n^{\text{bld,hig}}(\boldsymbol{\xi}_n)$ by introducing the auxiliary variables $P_{n,t}^{\text{bld}} := \sum_i P_{n,i,t}^{\text{hvac}}$ to (2), which represents the aggregate power of multiple zones at the building-level interface, as expressed in a compact form as below:

$$\mathbb{U}_n^{\text{bld,hig}}(\xi_n) := \left\{ \left[ \mathbf{T}_n^{\text{in}}; \mathbf{P}_n^{\text{hvac}}; \mathbf{P}_n^{\text{bld}} \right] \middle| \begin{array}{c} \mathbf{A}_n^{\text{eq1}}\mathbf{T}_n^{\text{in}} + \mathbf{A}_n^{\text{eq2}}\mathbf{P}_n^{\text{hvac}} + \mathbf{A}_n^{\text{eq3}}\xi_n = \mathbf{b}_n^{\text{eq}} \\ \mathbf{A}_n^{\text{ieq1}}\mathbf{T}_n^{\text{in}} \le \mathbf{b}_n^{\text{ieq1}}, \mathbf{A}_n^{\text{ieq2}}\mathbf{P}_n^{\text{hvac}} \le \mathbf{b}_n^{\text{ieq2}} \\ \mathbf{P}_n^{\text{bld}} = \mathbf{\Gamma}_n^{\text{ztb}}\mathbf{P}_n^{\text{hvac}} \end{array} \right\} \quad (4)$$

where $\mathbf{P}_n^{\text{bld}}:=\left[P_{n,1}^{\text{bld}};\ldots;P_{n,N^T}^{\text{bld}}\right]\in\mathbb{R}^{N^T}$; $\mathbf{\Gamma}_n^{\text{ztb}}\in\mathbb{R}^{N^{\text{T}}\times(N_n^{\text{I}}N^{\text{T}})}$ is the mapping matrix from zone-level power space to building-level power space. Due to $P_{n,t}^{\text{bld}}=\sum_i P_{n,i,t}^{\text{hvac}}$, the $r-$th row of $\mathbf{\Gamma}_n^{\text{ztb}}$, denoted by $\mathbf{\Gamma}_{n,r}^{\text{ztb}}$, is expressed as below:

$$\mathbf{\Gamma}_{n,r}^{\text{ztb}}:=\left[\mathbf{0}_r^{\text{ztb,lhs}},\mathbf{1}_r^{\text{ztb}},\mathbf{0}_r^{\text{ztb,rhs}}\right] \tag{5}$$

where $\mathbf{0}_r^{\text{ztb,lhs}}\in\mathbb{R}^{1\times(r-1)N^{\text{T}}}$ is the left-hand side all-zeros matrix, $\mathbf{1}_r^{\text{ztb}}\in\mathbb{R}^{1\times N^{\text{T}}}$ is all-ones matrix, and $\mathbf{0}_r^{\text{ztb,rhs}}\in\mathbb{R}^{(N^{\text{T}}-r)\times N^{\text{T}}}$ is the right-hand side all-zeros matrix.

By applying the above dimensional-lifting procedure, $\mathbb{P}^{\text{agg}}(\boldsymbol{\xi}_n)$ is reformulated as a decoupled Minkowski sum of the building-level aggregate powers $\mathbf{P}_n^{\text{bld}}$, as shown in (6). This reformulation confines all multi-zone coupling to $\mathbb{U}_n^{\text{bld,hig}}(\boldsymbol{\xi}_n)$, thereby markedly reducing the computational burden of the subsequent inter-building Minkowski-sum calculation.

$$\mathbb{P}^{\text{agg}}(\xi_n)=\left\{\mathbf{P}^{\text{agg}}\left|P_t^{\text{agg}}=\sum_n P_{n,t}^{\text{bld}},\underbrace{P_{n,t}^{\text{bld}}\in\mathbb{U}_n^{\text{bld,hig}}(\xi_n)}_{\text{decoupled sets}}\right.\right\} \tag{6}$$

However, we note that the resulting polytope $\mathbb{U}_n^{\text{bld,hig}}(\boldsymbol{\xi}_n)$ is still difficult to work with. Because it contains many equality constraints, it is not a full dimensional polytope, which triggers dimensional collapse in the subsequent inner approximations [7]. Moreover, it is affected by uncertain parameters $\boldsymbol{\xi}_n$ which propagate through many coupled variables. As a result, the impacts of $\boldsymbol{\xi}_n$ on $\mathbf{P}_n^{\text{bld}}$ are hard to bound, which compromises aggregation reliability. To resolve both issues, **Steps 2** and **Step 3** will remove the equality constraints and recast $\mathbb{U}_n^{\text{bld,hig}}(\boldsymbol{\xi}_n)$ into a concise polytope that is simultaneously low- and full-dimensional, thereby streamlining all later computations.

**Step 2** (Removing temperature state variables): To remove the equality constraints imposed by the thermal dynamics, we perform Gaussian elimination. Because the temperature evolution in (1.a) is driven by the controllable input $\mathbf{P}_n^{\text{hvac}}$, the system matrix $\mathbf{A}_n^{\text{eq1}}$ in (2) is invertible. Leveraging this invertibility, we rewrite the thermal dynamics as (7) and substitute (7) into $\mathbb{U}_n^{\text{bld,hig}}(\boldsymbol{\xi}_n)$. This substitution yields the medium-dimensional polytope $\mathbb{U}_n^{\text{bld,mid}}(\boldsymbol{\xi}_n)$ as shown in (8), where the coefficient matrices $\mathbf{A}_n^{\text{ieq3a}}$, $\mathbf{A}_n^{\text{ieq3b}}$, and $\mathbf{b}_n^{\text{ieq3}}$ are obtained by combining (4) with the reformulated equation (7).

$$\mathbf{T}_n^{\text{in}}=(\mathbf{A}_n^{\text{eq1}})^{-1}(\mathbf{b}_n^{\text{eq}}-\mathbf{A}_n^{\text{eq2}}\mathbf{P}_n^{\text{hvac}}-\mathbf{A}_n^{\text{eq3}}\xi_n) \tag{7}$$

$$\mathbb{U}_n^{\text{bld,mid}}(\xi_n):=\left\{\left[\mathbf{P}_n^{\text{hvac}};\mathbf{P}_n^{\text{bld}}\right]\left|\begin{array}{c}\mathbf{A}_n^{\text{ieq3a}}\mathbf{P}_n^{\text{hvac}}+\mathbf{A}_n^{\text{ieq3b}}\xi_n\le\mathbf{b}_n^{\text{ieq3}}\\ \mathbf{P}_n^{\text{bld}}=\mathbf{\Gamma}_n^{\text{ztb}}\mathbf{P}_n^{\text{hvac}}\end{array}\right.\right\} \tag{8}$$

**Step 3** (Projection onto a full-dimensional building-level polytope): To further remove the equality constraints introduced by the power-coupling equation (the second row of (8)), we propose a tailored coordinate transformation to handle. The following proposition formalizes the procedure:

**Proposition 1** (Projection via coordinate transformation). Let $\mathbb{P}_n^{\text{bld,mid}}(\boldsymbol{\xi}_n)$ and $\mathbb{P}_n^{\text{bld,low}}(\boldsymbol{\xi}_n)$ be the two polytopes specified in (9.a) and (9.b) where $\mathbf{\Gamma}_n^{\text{ztb}}$ is a full-row-rank matrix. Then, the relation in (9.c) holds. Here, $\mathbf{P}_n^{\text{aux}}$ is the auxiliary power vector with $\dim(\mathbf{P}_n^{\text{aux}})=\dim\left(\mathbf{P}_n^{\text{hvac}}\right)-\text{rank}(\mathbf{\Gamma}_n^{\text{ztb}})$ ; $\mathbf{\Gamma}_n^{\text{ct}}$ is a maximal linearly independent group of $\left[\mathbf{I}_n^{\text{hvac}};\mathbf{\Gamma}_n^{\text{ztb}}\right]$ that contains $\mathbf{\Gamma}_n^{\text{ztb}}$; $\mathbf{I}_n^{\text{hvac}}\in\mathbb{R}^{(N_n^{\text{I}}N^{\text{T}})\times(N_n^{\text{I}}N^{\text{T}})}$ is the identity matrix.

$$\mathbb{P}_n^{\text{bld,mid}}(\xi_n):=\left\{\mathbf{P}_n^{\text{bld}}\left|\begin{array}{c}\exists\mathbf{P}_n^{\text{hvac}},\text{such that:}\\ \mathbf{A}_n^{\text{ieq3a}}\mathbf{P}_n^{\text{hvac}}+\mathbf{A}_n^{\text{ieq3b}}\xi_n\le\mathbf{b}_n^{\text{ieq3}}\\ \mathbf{P}_n^{\text{bld}}=\mathbf{\Gamma}_n^{\text{ztb}}\mathbf{P}_n^{\text{hvac}}\end{array}\right.\right\} \tag{9.a}$$

$$\mathbb{P}_n^{\text{bld,low}}(\xi_n):=\left\{\mathbf{P}_n^{\text{bld}}\left|\begin{array}{c}\exists\mathbf{P}_n^{\text{aux}},\text{such that:}\\ \mathbf{A}_n^{\text{ieq3a}}(\mathbf{\Gamma}_n^{\text{ct}})^{-1}[\mathbf{P}_n^{\text{aux}};\mathbf{P}_n^{\text{bld}}]+\mathbf{A}_n^{\text{ieq3b}}\xi_n\le\mathbf{b}_n^{\text{ieq3}}\end{array}\right.\right\} \tag{9.b}$$

$$\mathbb{P}_n^{\text{bld,mid}}(\xi_n)=\mathbb{P}_n^{\text{bld,low}}(\xi_n) \tag{9.c}$$

*Proof*: See Appendix VI.A.

According to **Proposition 1**, we reformulate (8) into (10), where $[\mathbf{A}_n^{\text{ieq4a}},\mathbf{A}_n^{\text{ieq4b}}]=\mathbf{A}_n^{\text{ieq3a}}(\mathbf{\Gamma}_n^{\text{ct}})^{-1}$, $\mathbf{A}_n^{\text{ieq4c}}=\mathbf{A}_n^{\text{ieq3b}}$, and $\mathbf{b}_n^{\text{ieq4}}=\mathbf{b}_n^{\text{ieq3}}$.

$$\mathbb{U}_n^{\text{bld,low}}(\xi_n):=\left\{\left[\mathbf{P}_n^{\text{bld}};\mathbf{P}_n^{\text{aux}}\right]\middle|\mathbf{A}_n^{\text{ieq4a}}\mathbf{P}_n^{\text{bld}}+\mathbf{A}_n^{\text{ieq4b}}\mathbf{P}_n^{\text{aux}}+\mathbf{A}_n^{\text{ieq4c}}\xi_n\le\mathbf{b}_n^{\text{ieq4}}\right\} \tag{10}$$

After removing all equality constraints and redundant variables, we equivalently convert the high-dimensional set $\mathbb{U}_n^{\text{bld,hig}}(\boldsymbol{\xi}_n)$ into the low-dimensional building-level polytope $\mathbb{U}_n^{\text{bld,low}}(\boldsymbol{\xi}_n)$. Although this concise analytical expression (10) has fewer dimensions, we note that its projection onto the aggregate power $\mathbf{P}_n^{\text{bld}}$ is identical to that of the original set $\mathbb{U}_n^{\text{bld,hig}}(\boldsymbol{\xi}_n)$, so no flexibility is lost during the projection.

This reformulation offers two main advantages:

*1) Physical Insight:* The compact expression (10) clarifies how the zone-level uncertainties $\boldsymbol{\xi}_n$ propagate to the building-level interface $\mathbf{P}_n^{\text{bld}}$. It is obvious to see that the geometric structure of $\mathbb{U}_n^{\text{bld,low}}(\boldsymbol{\xi}_n)$ depend on extra auxiliary variables $\mathbf{P}_n^{\text{aux}}$ and forecast error uncertainties $\boldsymbol{\xi}_n$. The term $\mathbf{A}_n^{\text{ieq4b}}\mathbf{P}_{\text{n}}^{\text{aux}}$ captures the impact of the redundancy of $\mathbf{P}_n^{\text{hvac}}$ on $\mathbb{U}_n^{\text{bld,low}}(\boldsymbol{\xi}_n)$, while $\mathbf{A}_n^{\text{ieq4c}}\boldsymbol{\xi}_n$ reflects the influence of the inherent randomness of $\boldsymbol{\xi}_n$ on $\mathbb{U}_n^{\text{bld,low}}(\boldsymbol{\xi}_n)$. The coefficient matrices $\mathbf{A}_n^{\text{ieq4a}},\mathbf{A}_n^{\text{ieq4b}}$, and $\mathbf{A}_n^{\text{ieq4c}}$ together reveal that the propagation mechanism of $\boldsymbol{\xi}_n$ depends on both the original multi-zone coupling and the applied coordinate transformation.

*2) Mathematical Tractability:* The original polytope $\mathbb{U}_n^{\text{bld,hig}}(\boldsymbol{\xi}_n)$, replete with many equality constraints and redundant variables, is non-full-dimensional, hindering downstream aggregator-level computations. In contrast, $\mathbb{U}_n^{\text{bld,low}}(\boldsymbol{\xi}_n)$ is full-dimensional in the new coordination, offering desirable geometric properties. This structure enables us to efficiently calculate the optimal aggregate parameters under multi-zone coupling and uncertainty propagation, which will be detailed in Section III.D.

With $\mathbb{U}_n^{\text{bld,low}}(\boldsymbol{\xi}_n)$ in hand, we can now deal with large-scale inter-building aggregation problem at the aggregator level.

*C. Aggregator-Level Inner Approximation*

Although $\mathbb{U}_n^{\text{bld,low}}(\boldsymbol{\xi}_n)$ is lower-dimensional than the original $\mathbb{U}_n^{\text{bld,hig}}(\boldsymbol{\xi}_n)$, it still includes the auxiliary variable $\mathbf{P}_n^{\text{aux}}$ rather than being defined solely by the building-level aggregate power $\mathbf{P}_n^{\text{bld}}$. This creates a dimensional mismatch that makes existing polytope-based inner approximation approaches inapplicable, since these methods require matching dimensions between inner and target polytopes.

A natural solution would be to directly project $\mathbb{U}_n^{\text{bld,low}}(\boldsymbol{\xi}_n)$ onto the subspace spanned by $\mathbf{P}_n^{\text{bld}}$ (the projected set is denoted by $\mathbb{P}_n^{\text{bld}}(\boldsymbol{\xi}_n)$) and then approximate that projection $\mathbb{P}_n^{\text{bld}}(\boldsymbol{\xi}_n)$. However, $\mathbb{U}_n^{\text{bld,low}}(\boldsymbol{\xi}_n)$ couples multiple time periods, zones, and uncertainties, making exact projection NP-hard [13]. Therefore, traditional polytope-based approaches remain inapplicable due to their restrictive dimensional requirements.

To address this challenge, we propose a matrix-transformation technique that reformulates the inner approximation between different-dimensional polytopes as an affine polytope containment problem (Fig. 5). This approach avoids computing exact projections, which are both computationally expensive and unnecessary since they serve only as intermediate steps.

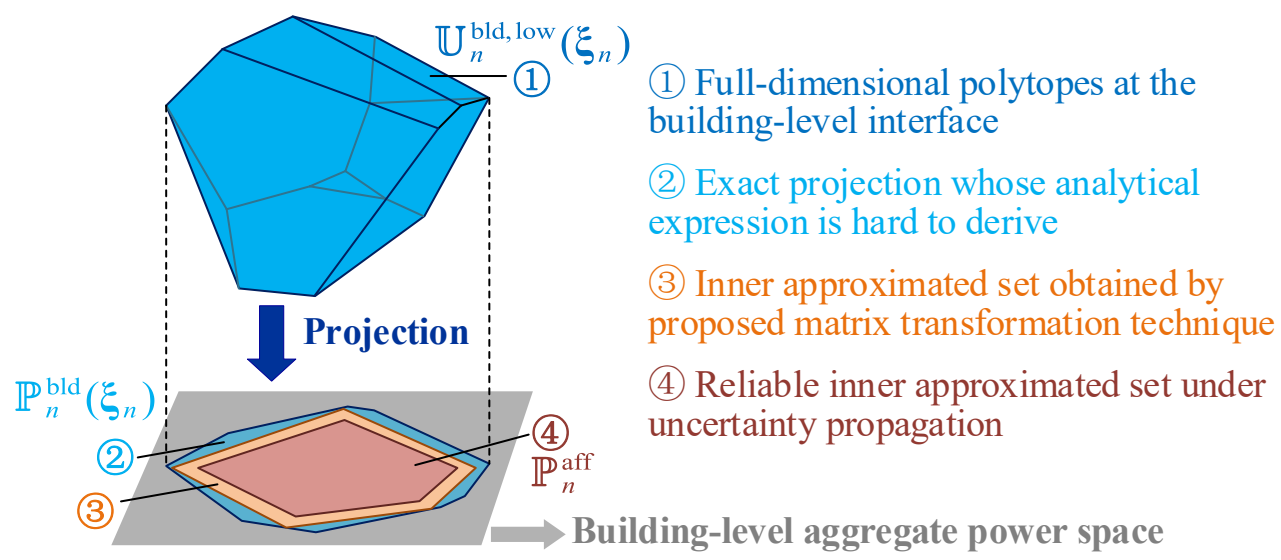

Fig. 5. Illustration of inner approximation.

Specifically, our approach involves three key steps to calculate the final aggregate set $\mathbb{P}^{\text{agg}}$ in closed form:

**Step 1** (Reformulation via matrix transformation): Although the explicit expression of $\mathbb{P}_n^{\text{bld}}(\boldsymbol{\xi}_n)$ is intractable, fortunately, there exists the following linear mapping which represents the projection relationship between $\mathbb{P}_n^{\text{bld}}(\boldsymbol{\xi}_n)$ and $\mathbb{U}_n^{\text{bld,low}}(\boldsymbol{\xi}_n)$:

$$\mathbf{P}_n^{\text{bld}} = \boldsymbol{\Gamma}_n^{\text{ltp}}\left[\mathbf{P}_n^{\text{bld}};\mathbf{P}_n^{\text{aux}}\right] \tag{11}$$

where $\boldsymbol{\Gamma}_n^{\text{ltp}} \in \mathbb{R}^{N^{\text{T}} \times N_n^{\text{I}} N^{\text{T}}}$ is the dimensional transformation matrix as defined in (12):

$$\boldsymbol{\Gamma}_n^{\text{ltp}} := [\mathbf{I}_n^{\text{ltp,lhs}}, \mathbf{0}_n^{\text{ltp,rhs}}] \tag{12}$$

where $\mathbf{I}_n^{\text{ltp,lhs}} \in \mathbb{R}^{N^{\text{T}} \times N^{\text{T}}}$ is the left-hand-side identity matrix and $\mathbf{0}_n^{\text{ltp,rhs}} \in \mathbb{R}^{N^{\text{T}} \times (N_n^{\text{I}} N^{\text{T}} - N^{\text{T}})}$ is the right-hand-side zero matrix.

Leveraging (11), we obtain an indirect yet explicit form of $\mathbb{P}_n^{\text{bld}}(\boldsymbol{\xi}_n)$ as below:

$$\mathbb{P}_n^{\text{bld}}(\boldsymbol{\xi}_n) = \boldsymbol{\Gamma}_n^{\text{ltp}} \mathbb{U}_n^{\text{bld,low}}(\boldsymbol{\xi}_n) \tag{13}$$

**Step 2** (Inner approximation for explicit expression): Drawing inspiration from established polytope-based methods [3], [8], [11], we approximate $\mathbb{P}_n^{\text{bld}}(\boldsymbol{\xi}_n)$ as the affine image of a chosen full-dimensional base set $\mathbb{P}^{\text{base}} \subseteq \mathbb{R}^{N^{\text{T}}}$. Specifically, by applying a linear transformation and translation, we obtain an explicit and accurate representation $\mathbb{P}_n^{\text{aff}}$, as shown in (14). This inner approximation offers two key advantages. First, $\mathbb{P}_n^{\text{aff}}$ admits a closed-form Minkowski sum, thereby facilitating large-scale aggregation (see (17)). Second, compared with other inner-approximation techniques, the affine mapping provides greater transformation flexibility and thus higher aggregation accuracy for the same base set [3].

$$\mathbb{P}_n^{\text{aff}} := \boldsymbol{\gamma}_n^{\text{aff}} + \boldsymbol{\Gamma}_n^{\text{aff}} \mathbb{P}^{\text{base}}, \mathbb{P}^{\text{base}} := \left\{\mathbf{P}^{\text{base}} \mid \mathbf{H}^{\text{base}} \mathbf{P}^{\text{base}} \leq \mathbf{h}^{\text{base}}\right\} \tag{14}$$

where $\boldsymbol{\Gamma}_n^{\text{aff}} \in \mathbb{R}^{N^{\text{T}} \times N^{\text{T}}}$ is a linear transformation matrix; $\boldsymbol{\gamma}_n^{\text{aff}} \in \mathbb{R}^{N^{\text{T}}}$ is a translation vector; both $\boldsymbol{\Gamma}_n^{\text{aff}}$ and $\boldsymbol{\gamma}_n^{\text{aff}}$ are considered as aggregate parameters that will be determined in Section III.D; the coefficient matrix $\boldsymbol{H}^{\text{base}}$ and vector $\boldsymbol{h}^{\text{base}}$ are predefined parameters, which can be specified via averaging the zone thermal parameters [3].

Additionally, to ensure $\mathbb{P}_n^{\text{aff}} \subseteq \mathbb{P}_n^{\text{bld}}(\boldsymbol{\xi}_n) = \boldsymbol{\Gamma}_n^{\text{ltp}} \mathbb{U}_n^{\text{bld,low}}(\boldsymbol{\xi}_n)$ with high probability $1-\varepsilon$ under the uncertainty of $\boldsymbol{\xi}_n$, we employ DRCCs to characterize the inner approximation as below:

$$\inf_{f(\xi_n)\in\mathbb{D}} \mathbf{Pr}\left\{\boldsymbol{\gamma}_n^{\text{aff}} + \boldsymbol{\Gamma}_n^{\text{aff}} \mathbb{P}^{\text{base}} \subseteq \boldsymbol{\Gamma}_n^{\text{ltp}} \mathbb{U}_n^{\text{bld,l}}(\xi_n)\right\} \geq 1-\varepsilon \tag{15}$$

where $\varepsilon$ is the allowable violation probability; $\mathbf{Pr}\{\cdot\}$ represents the probability measure; $f(\boldsymbol{\xi}_n)$ is the probability distribution function; the ambiguity set $\mathbb{D}$ is built as shown in (16):

$$\mathbb{D} := \left\{\mathbf{E}[\xi_n] = \boldsymbol{\mu}_n, \mathbf{E}\left[(\xi_n - \boldsymbol{\mu}_n)(\xi_n - \boldsymbol{\mu}_n)^{\text{T}}\right] = \boldsymbol{\sigma}_n\right\} \tag{16}$$

where $\mathbf{E}[\cdot]$ is the expectation operator; $\boldsymbol{\mu}_n$ is empirical mean vector; $\boldsymbol{\sigma}_n$ is empirical covariance matrix.

**Step 3** (Aggregation via closed-form calculation): Once the aggregate parameters, $\boldsymbol{\Gamma}_n^{\text{aff}}$ and $\boldsymbol{\gamma}_n^{\text{aff}}$, are obtained for each building, the final approximate aggregate set $\mathbb{P}^{\text{app}}$ can be calculated using the following closed-form expression [3]:

$$\mathbb{P}^{\text{app}} = \boldsymbol{\Gamma}^{\text{agg}} \mathbb{P}^{base} + \boldsymbol{\gamma}^{\text{agg}} \subseteq \mathbb{P}^{\text{agg}}(\xi_n) \tag{17}$$

where $\boldsymbol{\Gamma}^{\text{agg}} := \sum_n(\boldsymbol{\Gamma}_n^{\text{aff}})$ and $\boldsymbol{\gamma}^{\text{agg}} := \sum_n(\boldsymbol{\gamma}_n^{\text{aff}})$.

**Remark 1** (Comparison to existing polytope-based approaches). Existing polytope-based approaches [1], [3], [6], [7], [8], [11] approximate each feasibility set $\mathbb{U}_n^{\text{DER}}$ by constructing an inner approximation of the form $\mathbb{P}_n^{\text{aff}} = \boldsymbol{\gamma}_n^{\text{aff}} + \boldsymbol{\Gamma}_n^{\text{aff}} \mathbb{P}^{\text{base}} \subseteq \mathbb{U}_n^{\text{DER}}$, and then compute their Minkowski sum via (17). A key assumption is that $\mathbb{P}_n^{\text{aff}}$ and $\mathbb{U}_n^{\text{DER}}$ share the same dimensionality. However, applying these techniques directly to the high-dimensional, multi-zone-coupled sets $\mathbb{U}_n^{\text{bld,low}}(\boldsymbol{\xi}_n)$ does not guarantee that $\mathbb{P}_n^{\text{aff}}$ lies within the power subspace of $\mathbb{U}_n^{\text{bld,low}}(\boldsymbol{\xi}_n)$. If the inner approximation spans non-power dimensions, the subsequent Minkowski-sum formula (17) fails since it sums only power components. As a result, existing polytope-based approaches no longer work in this scenario.

By contrast, our approach introduces a dimensional transformation matrix $\boldsymbol{\Gamma}_n^{\text{ltp}}$ as shown in (15) to flexibly match the dimensions of any polytope. This lets us directly approximate the projection of $\mathbb{U}_n^{\text{bld,low}}(\boldsymbol{\xi}_n)$ onto the aggregate power space, ensuring the resulting polytope resides purely in power dimensions and remains summable. Moreover, when $\boldsymbol{\Gamma}_n^{\text{ltp}}$ is set to the identity matrix, our approach reverts to the existing ones. Therefore, our approach effectively generalizes existing approaches to multi-zone-coupled HVAC loads.

A digestible case has been given in Appendix VI.E to illustrate the advantages of the proposed matrix transformation technique over traditional inner approximations.

*D. Aggregate Parameter Determination*

To determine the optimal aggregate parameters $\boldsymbol{\Gamma}^{\text{agg}}$ and $\boldsymbol{\gamma}^{\text{agg}}$ in (17), we proceed in two steps. First, we recast the inner-approximation condition as a set of linear constraints to ensure computational tractability. Second, we formulate a linear program to choose the optimal $\boldsymbol{\Gamma}^{\text{agg}}$ and $\boldsymbol{\gamma}^{\text{agg}}$. The resulting parameters ensure that $\mathbb{P}^{\text{app}}$ is the largest-volume inner approximation of $\mathbb{P}^{\text{agg}}(\boldsymbol{\xi}_n)$ under uncertain $\boldsymbol{\xi}_n$, thereby preserving aggregation accuracy and reliability.

*1) Tractable Reformulation*: Analytically expressing the polytope inclusion relationship $\boldsymbol{\Gamma}^{\text{agg}}\mathbb{P}^{\text{base}} + \boldsymbol{\gamma}^{\text{agg}} \subseteq \mathbb{P}^{\text{agg}}(\boldsymbol{\xi}_n)$ is computationally challenging. Fortunately, both $\mathbb{P}^{\text{base}}$ and $\mathbb{U}_n^{\text{bld,low}}(\boldsymbol{\xi}_n)$ are full-dimensional as demonstrated in Section III.B, so we can invoke Farkas' Lemma [34], [35] for such sets to derive **Proposition 2**, as stated below.

**Proposition 2** (Constraint linearization). It holds that $\boldsymbol{\Gamma}^{\text{agg}}\mathbb{P}^{\text{base}} + \boldsymbol{\gamma}^{\text{agg}} \subseteq \mathbb{P}^{\text{agg}}(\boldsymbol{\xi}_n)$ if there exist decision variables $\boldsymbol{\Gamma}_n^{\text{aff}} \in \mathbb{R}^{N^{\text{T}}\times N^{\text{T}}}$, $\boldsymbol{\gamma}_n^{\text{aff}} \in \mathbb{R}^{N^{\text{T}}}$, $\mathbf{G}_n^{\text{aux}} \in \mathbb{R}^{N_n^{\text{I}}N^{\text{T}}\times N^{\text{T}}}$, $\boldsymbol{\Lambda}_n^{\text{aux}} \in \mathbb{R}_+^{\dim(\mathbf{b}_n^{\text{ieq4}})\times\dim(\mathbf{h}^{\text{base}})}$, and $\boldsymbol{\beta}_n^{\text{aux}} \in \mathbb{R}^{N_n^{\text{I}}N^{\text{T}}}$ such that:

$$\boldsymbol{\Gamma}^{\text{agg}} = \sum_n \boldsymbol{\Gamma}_n^{\text{aff}}, \boldsymbol{\gamma}^{agg} = \sum_n \boldsymbol{\gamma}_n^{aff} \tag{18.a}$$

$$\boldsymbol{\Gamma}_n^{\text{aff}} = \boldsymbol{\Gamma}_n^{\text{ltp}}\mathbf{G}_n^{\text{aux}}, -\boldsymbol{\gamma}_n^{\text{aff}} = \boldsymbol{\Gamma}_n^{\text{ltp}}\boldsymbol{\beta}_n^{\text{aux}} \tag{18.b}$$

$$\boldsymbol{\Lambda}_n^{\text{aux}}\mathbf{H}^{\text{base}} = [\mathbf{A}_n^{\text{ieq4a}}, \mathbf{A}_n^{\text{ieq4b}}]\mathbf{G}_n^{\text{aux}} \tag{18.c}$$

$$\boldsymbol{\Lambda}_n^{\text{aux}}\mathbf{h}^{\text{base}} + \mathbf{A}_n^{\text{ieq4c}}\boldsymbol{\xi}_n \leq \mathbf{b}_n^{\text{ieq4}} + [\mathbf{A}_n^{\text{ieq4a}}, \mathbf{A}_n^{\text{ieq4b}}]\boldsymbol{\beta}_n^{\text{aux}} \tag{18.d}$$

*Proof.* See Appendix VI.B.

From **Proposition 2**, we recast (15) as below to facilitate linear expression:

$$\begin{cases} \boldsymbol{\Gamma}^{\text{agg}} = \sum_n \boldsymbol{\Gamma}_n^{\text{aff}}, \boldsymbol{\gamma}^{\text{agg}} = \sum_n \boldsymbol{\gamma}_n^{\text{aff}} \\ \boldsymbol{\Gamma}_n^{\text{aff}} = \boldsymbol{\Gamma}_n^{\text{ltp}}\mathbf{G}_n^{\text{aux}}, -\boldsymbol{\gamma}_n^{\text{aff}} = \boldsymbol{\Gamma}_n^{\text{ltp}}\boldsymbol{\beta}_n^{\text{aux}}, \boldsymbol{\Lambda}_n^{\text{aux}}\mathbf{H}^{\text{base}} = [\mathbf{A}_n^{\text{ieq4a}}, \mathbf{A}_n^{\text{ieq4b}}]\mathbf{G}_n^{\text{aux}} \\ \inf_{f(\boldsymbol{\xi}_n)\in\mathbb{D}} \mathbf{Pr}\left\{\boldsymbol{\Lambda}_n^{\text{aux}}\mathbf{h}^{\text{base}} + \mathbf{A}_n^{\text{ieq4c}}\boldsymbol{\xi}_n \leq \mathbf{b}_n^{\text{ieq4}} + [\mathbf{A}_n^{\text{ieq4a}}, \mathbf{A}_n^{\text{ieq4b}}]\boldsymbol{\beta}_n^{\text{aux}}\right\} \geq 1-\varepsilon \end{cases} \tag{19}$$

Thanks to the full-dimensional nature of both $\mathbb{P}^{\text{base}}$ and $\mathbb{U}_n^{\text{bld,low}}(\boldsymbol{\xi}_n)$, the third line in (19) involves only half-space inequalities and no equalities. Hence, we can apply the Bonferroni approximation [36] to further recast the third line in (19) into the following tractable linear form:

$$\begin{aligned} &\boldsymbol{\Lambda}_n^{\text{aux}}\mathbf{h}^{\text{base}} + \mathbf{A}_n^{\text{ieq4c}}\boldsymbol{\mu}_n + \sqrt{(1-\varepsilon)/\varepsilon}\sqrt{\mathbf{A}_n^{\text{ieq4c}}\boldsymbol{\sigma}_n\left(\mathbf{A}_n^{\text{ieq4c}}\right)^{\text{T}}} \\ &\leq \mathbf{b}_n^{\text{ieq4}} + [\mathbf{A}_n^{\text{ieq4a}}, \mathbf{A}_n^{\text{ieq4b}}]\boldsymbol{\beta}_n^{\text{aux}} \end{aligned} \tag{20}$$

In formulation (20), $\varepsilon$ defines the trade-off between robustness and conservatism. Decreasing $\varepsilon$ improves robustness to the uncertainty $\boldsymbol{\xi}_n$, but simultaneously tightens the approximation, making it more conservative. Hence, $\varepsilon$ must be selected carefully—enough to safeguard against the observed uncertainty, yet not so small that it unduly restricts flexibility. The proposition that follows offers a principled rule for choosing $\varepsilon$, preventing excessive conservatism while preserving the robustness required for reliable aggregation.

**Proposition 3** (Data-driven $\varepsilon$-selection strategy). Let $\mathbb{X}_n$ be a collection of historical samples for building $n$, with empirical mean $\boldsymbol{\mu}_n$ and covariance $\boldsymbol{\sigma}_n$. For each sample $\boldsymbol{\xi}_n \in \mathbb{X}_n$, we define its robustness index $\Upsilon(\boldsymbol{\xi}_n) = \max_r \frac{\mathbf{A}_{n,r}^{\text{ieq4c}}(\boldsymbol{\xi}_n-\boldsymbol{\mu}_n)}{\sqrt{\mathbf{A}_{n,r}^{\text{ieq4c}}\boldsymbol{\sigma}_n\left(\mathbf{A}_{n,r}^{\text{ieq4c}}\right)^{\text{T}}}}$ where $r$ runs over the rows of $\mathbf{A}_{n,r}^{\text{ieq4c}}$. Let $\varepsilon^* = \frac{1}{1+\left(\max_{\boldsymbol{\xi}_n\in\mathbb{X}_n}\Upsilon(\boldsymbol{\xi}_n)\right)^2}$. Then choosing $\varepsilon = \varepsilon^*$ is the largest risk level that guarantees $\boldsymbol{\Lambda}_n^{\text{aux}}\mathbf{h}^{\text{base}} + \mathbf{A}_n^{\text{ieq4c}}\boldsymbol{\xi}_n \leq \mathbf{b}_n^{\text{ieq4}} + [\mathbf{A}_n^{\text{ieq4a}}, \mathbf{A}_n^{\text{ieq4b}}]\boldsymbol{\beta}_n^{\text{aux}}, \forall\boldsymbol{\xi}_n \in \mathbb{X}_n$.

*Proof.* See Appendix VI.C.

*2) Optimization Problem*: With the tractable reformulation in hand, we formulate the optimization problem (21.a)-(21.e) to select the optimal aggregate parameters $\boldsymbol{\Gamma}^{\text{agg}}$ and $\boldsymbol{\gamma}^{\text{agg}}$. Since the volume of $\mathbb{P}^{\text{app}}$ scales with the trace of $\boldsymbol{\Gamma}^{\text{agg}}$ [3], we maximize $\text{tr}(\boldsymbol{\Gamma}^{\text{agg}})$ so that $\mathbb{P}^{\text{app}} = \boldsymbol{\Gamma}^{\text{agg}}\mathbb{P}^{base} + \boldsymbol{\gamma}^{\text{agg}}$ is the largest-volume subset of $\mathbb{P}^{\text{agg}}(\boldsymbol{\xi}_n)$, thereby maximizing aggregation accuracy.

$$\max \text{tr}\left(\boldsymbol{\Gamma}^{\text{agg}}\right) \tag{21.a}$$

$$\text{s.t. } \boldsymbol{\Gamma}^{\text{agg}} = \sum_n \boldsymbol{\Gamma}_n^{\text{aff}}, \boldsymbol{\gamma}^{\text{agg}} = \sum_n \boldsymbol{\gamma}_n^{\text{aff}} \tag{21.b}$$

$$\boldsymbol{\Gamma}_n^{\text{aff}} = \boldsymbol{\Gamma}_n^{\text{ltp}}\mathbf{G}_n^{\text{aux}}, -\boldsymbol{\gamma}_n^{\text{aff}} = \boldsymbol{\Gamma}_n^{\text{ltp}}\boldsymbol{\beta}_n^{\text{aux}} \tag{21.c}$$

$$\boldsymbol{\Lambda}_n^{\text{aux}}\mathbf{H}^{\text{base}} = [\mathbf{A}_n^{\text{ieq4a}}, \mathbf{A}_n^{\text{ieq4b}}]\mathbf{G}_n^{\text{aux}} \tag{21.d}$$

$$\begin{aligned} &\boldsymbol{\Lambda}_n^{\text{aux}}\mathbf{h}^{\text{base}} + \mathbf{A}_n^{\text{ieq4c}}\boldsymbol{\mu}_n + \sqrt{(1-\varepsilon)/\varepsilon}\sqrt{\mathbf{A}_n^{\text{ieq4c}}\boldsymbol{\sigma}_n\left(\mathbf{A}_n^{\text{ieq4c}}\right)^{\text{T}}} \\ &\leq \mathbf{b}_n^{\text{ieq4}} + [\mathbf{A}_n^{\text{ieq4a}}, \mathbf{A}_n^{\text{ieq4b}}]\boldsymbol{\beta}_n^{\text{aux}} \end{aligned} \tag{21.e}$$

**Remark 2** (Computational complexity of our approach). We note that the inter-building decoupling renders the optimization problem (21.a)-(21.e) *block-separable* with respect to each building's decision variables. Hence, the global problem can decompose into $N^{\text{B}}$ independent linear-programming subproblems (see (22.a)-(22.d)), one for each building. These subproblems can be solved in parallel and then recombined via $\boldsymbol{\Gamma}^{\text{agg}} = \sum_n\left(\boldsymbol{\Gamma}_n^{\text{aff}}\right)$ and $\boldsymbol{\gamma}^{\text{agg}} = \sum_n\left(\boldsymbol{\gamma}_n^{\text{aff}}\right)$ to recover the optimal solution of the original formulation (21.a)-(21.e). This block-separable, parallelizable structure significantly accelerates computation, making the proposed aggregation framework well suited for large-scale implementation.

$$\max \text{tr}\left(\boldsymbol{\Gamma}_n^{\text{aff}}\right) \tag{22.a}$$

$$\text{s.t. } \boldsymbol{\Gamma}_n^{\text{aff}} = \boldsymbol{\Gamma}_n^{\text{ltp}}\mathbf{G}_n^{\text{aux}}, -\boldsymbol{\gamma}_n^{\text{aff}} = \boldsymbol{\Gamma}_n^{\text{ltp}}\boldsymbol{\beta}_n^{\text{aux}} \tag{22.b}$$

$$\boldsymbol{\Lambda}_n^{\text{aux}}\mathbf{H}^{\text{base}}=[\mathbf{A}_n^{\text{ieq4a}},\mathbf{A}_n^{\text{ieq4b}}]\mathbf{G}_n^{\text{aux}} \tag{22.c}$$

$$\begin{aligned}&\boldsymbol{\Lambda}_n^{\text{aux}}\mathbf{h}^{\text{base}}+\mathbf{A}_n^{\text{ieq4c}}\boldsymbol{\mu}_n+\sqrt{(1-\varepsilon)/\varepsilon}\sqrt{\mathbf{A}_n^{\text{ieq4c}}\boldsymbol{\sigma}_n\left(\mathbf{A}_n^{\text{ieq4c}}\right)^{\text{T}}}\\&\quad\le\mathbf{b}_n^{\text{ieq4}}+[\mathbf{A}_n^{\text{ieq4a}},\mathbf{A}_n^{\text{ieq4b}}]\boldsymbol{\beta}_n^{\text{aux}}\end{aligned} \tag{22.d}$$

### *E. Fast and Feasible Disaggregation Strategy*

After completing aggregation, the aggregator submits the approximate set $\mathbb{P}^{\text{app}}$ to the power system and obtains the scheduled aggregate power $\mathbf{P}^{\text{agg},*}\in\mathbb{P}^{\text{app}}$ (where the superscript * denotes scheduled values of decision variables). To implement this profile, the aggregator should decompose $\mathbf{P}^{\text{agg},*}$ into individual HVAC schedules, i.e., $\mathbf{P}^{\text{agg},*}=\sum_n\mathbf{P}_n^{\text{bld},*}$, $\mathbf{P}_n^{\text{bld},*}=\sum_i\mathbf{P}_{n,i}^{\text{hvac},*}$ with each trajectory lying within its feasible region. This can be achieved by solving an optimization problem directly, but its size grows with the number of buildings. Leveraging our aggregation framework, we therefore propose a corresponding two-level disaggregation strategy that ensures both speed and feasibility, as outlined below:

*1) Building-Level Disaggregation*: The **Proposition 4** presents a closed-form method to split any aggregate power point $\mathbf{P}^{\text{agg},*}$ into building-level allocations, i.e., $\mathbf{P}^{\text{agg},*}=\sum_n\mathbf{P}_n^{\text{bld},*}$. Specifically, for a given $\mathbf{P}^{\text{agg},*}$, one can compute each $\mathbf{P}_n^{\text{bld},*}$ from (23) so that every building's allocation lies within its feasible region and that the aggregate equals the scheduled profile.

**Proposition 4** (Building-Level disaggregation). For any scheduled aggregate power profile $\mathbf{P}^{\text{agg},*}\in\mathbb{P}^{\text{app}}$, there exist building-level power profiles $\mathbf{P}_n^{\text{bld},*}$ from (23) such that both $\mathbf{P}^{\text{agg},*}=\sum_n\mathbf{P}_n^{\text{bld},*}$ and $\mathbf{P}_n^{\text{bld},*}\in\boldsymbol{\Gamma}_n^{\text{ltp}}\mathbb{U}_n^{\text{bld,low}}(\boldsymbol{\xi}_n)$ hold.

$$\mathbf{P}_n^{\text{bld},*}=\boldsymbol{\gamma}_n^{\text{aff}}+\boldsymbol{\Gamma}_n^{\text{aff}}\left(\boldsymbol{\Gamma}^{\text{agg}}\right)^{-1}\left(\mathbf{P}^{\text{agg},*}-\boldsymbol{\gamma}^{\text{agg}}\right) \tag{23}$$

*Proof*. See Appendix VI.D.

*2) Zone-Level Disaggregation*: Once the building-level power $\mathbf{P}_n^{\text{bld},*}$ is set, we allocate it to each HVAC unit in building $n$. For each building, we solve the optimization problem (24.a) -(24.e) to compute zone power $\mathbf{P}_{n,i}^{\text{hvac},*}$ and indoor temperatures.

$$\min\sum_i\sum_t\left|S_{n,i,t}^{\text{nt}}-S_{n,t}^{\text{aux}}\right| \tag{24.a}$$

$$\text{s.t. } S_{n,i,t}^{\text{nt}}=(T_{n,i,t}^{\text{in}}-T_{n,i}^{\text{set}})/\beta_{n,i}^{\text{set}} \tag{24.b}$$

$$\mathbf{A}_n^{\text{eq1}}\mathbf{T}_n^{\text{in}}+\mathbf{A}_n^{\text{eq2}}\mathbf{P}_n^{\text{hvac}}+\mathbf{A}_n^{\text{eq3}}\boldsymbol{\xi}_n=\mathbf{b}_n^{\text{eq}} \tag{24.c}$$

$$\mathbf{A}_n^{\text{ieq1}}\mathbf{T}_n^{\text{in}}\le\mathbf{b}_n^{\text{ieq1}},\mathbf{A}_n^{\text{ieq2}}\mathbf{P}_n^{\text{hvac}}\le\mathbf{b}_n^{\text{ieq2}} \tag{24.d}$$

$$\mathbf{P}_n^{\text{bld},*}=\boldsymbol{\Gamma}_n^{\text{ztb}}\mathbf{P}_n^{\text{hvac}} \tag{24.e}$$

Here, we introduce the auxiliary variables $S_{n,t}^{\text{aux}}$ for each building and the normalized temperatures $S_{n,i,t}^{\text{nt}}$ for each zone. To accommodate distinct thermal dynamics, we map raw indoor temperatures $T_{n,i,t}^{\text{in}}$ to the interval $[-1,1]$ via constraints (24.b) and (1.b). This unified scale enables direct comparison of comfort across zones. The objective (24.a) minimizes deviations among these normalized temperatures, promoting uniform comfort. Since $\mathbf{P}_n^{\text{bld},*}\in\boldsymbol{\Gamma}_n^{\text{ltp}}\mathbb{U}_n^{\text{bld,low}}(\boldsymbol{\xi}_n)$, constraints (24.a)-(24.e) always admit a feasible solution. Therefore, the zonal disaggregation problem is guaranteed feasible.

**Remark 3** (Disaggregation strategy). The proposed two-level disaggregation approach balances three aspects:

*a) Computational Efficiency*: At the building level, algebraic rules (23) allocate scheduled power to each building using closed-form expressions. This avoids large-scale optimization problems whose complexity grows with building count, enabling near-instant computation for large systems.

*b) Fairness within Buildings*: At the zone level, smaller optimization problems (24.a)-(24.e) distribute each building's allocation among its zones. While more computationally demanding, these problems can be solved independently for each building and executed in parallel, maintaining low overall runtime. The objective minimizes deviations in normalized zone temperatures to ensure consistent comfort across zones.

*c) Operational Feasibility*: Since our aggregate flexibility set uses inner approximation, the disaggregation process always produces feasible solutions.

We acknowledge that the current strategy does not address all real-world complexities. Future work could extend the proposed framework to handle heterogeneous building response patterns and inter-building fairness considerations, though these extensions fall outside this paper's main scope.

## IV. CASE STUDIES

### *A. Simulation Setup*

*1) Simulation Problem*: To compare the quality of different aggregation approaches, we formulate the following economic cost problem at the aggregator level:

$$\min_{\mathbf{P}^{\text{agg}}\in\mathbb{P}}\sum_t c_t P_t^{\text{agg}}\Delta t \tag{25}$$

where $c_t$ (EUR/MWh) is the energy price; $\Delta t$ is the period length; $\mathbb{P}$ is the aggregate flexibility set. To quantify how much flexibility the inner approximation leaves unused under varying price profiles, we define the unused potential ratio (UPR) as:

$$\delta^{\text{UPR}}=\frac{z^{\text{app}}-z^{\text{exa}}}{z^{\text{noctr}}-z^{\text{exa}}}\times100\% \tag{26}$$

where $z^{\text{app}}$ is the optimal cost using the approximate flexibility set, $z^{\text{exa}}$ is the cost using the exact Minkowski-sum region (no approximation), and $z^{\text{noctr}}$ is the cost in a baseline "no-control" scenario, i.e., setting $T_{n,i,t}^{\text{in}}=T_{n,i}^{\text{set}}$ for each period. A $\delta^{\text{UPR}}$ near 0% means the approximation yields virtually the same cost as the exact region, whereas a $\delta^{\text{UPR}}$ close to 100% indicates substantial flexibility remains untapped.

*2) Simulation Setting*: We conduct our experiments on a 31-day dataset of day-ahead, hourly energy prices (July 1–31, 2024) drawn from the Spanish power system [37]. There are 100 buildings for testing with parameters given in TABLE II. The parameters are specified by its fixed value or the interval used for random sampling. $\alpha$ refers to the heterogeneity of HVAC parameters, and is set as 1 unless specified. All simulations are with an Intel i9-14900 CPU 2.20GHz and 64 GB RAM, using GUROBI 12.0.1 [38] for calculation.

TABLE II
HVAC PARAMETERS IN CASE STUDIES

| Item | Value | Unit | Item | Value | Unit |
|---|---|---|---|---|---|
| $N^{\mathrm{T}}$ | 24 | / | $\Delta t$ | 1 | hour |
| $N^{\mathrm{B}}$ | 100 | / | $N^{\mathrm{I}}$ | 2-8 | / |
| $R^{\mathrm{in}}$ | $4 \pm 2\alpha$ | ℃/kW | $R^{\mathrm{out}}$ | $2 \pm 1\alpha$ | ℃/kW |
| $C^{\mathrm{in}}$ | $5 \pm 3\alpha$ | kWh/℃ | $\eta^{\mathrm{hvac}}$ | $3 \pm 1\alpha$ | p.u. |
| $T^{\mathrm{set}}$ | $20 \pm 2\alpha$ | ℃ | $\beta^{\mathrm{set}}$ | $2 \pm 1\alpha$ | ℃ |
| $P^{\mathrm{hvac,min}}$ | $0.5 \pm 0.2\alpha$ | kW | $P^{\mathrm{hvac,max}}$ | $7 \pm 3\alpha$ | kW |

### *B. Validation of Aggregation Accuracy*

In this subsection, we evaluate the aggregation accuracy of the proposed approach under deterministic conditions; uncertainty will be addressed in the next subsection.

*1) A Toy Illustrative Example of Inner Approximation*: We begin with a three-dimensional illustrative example ($N^{\mathrm{T}} = 3$) to validate the proposed building-level inner approximation. A four-zone-coupled HVAC load in a building, with parameters listed in TABLE II, is used to compare the following five approaches for approximating $\mathbb{P}^{\mathrm{bld}}$:

**M0**: Exact set obtained with Fourier–Motzkin elimination. Note that while the algorithm yields a precise flexibility set, it suffers from the curse of dimensionality in high-dimensional implementation and is therefore used only as a benchmark for evaluation in the toy example [39].

**M1 (Proposed)**: Approximate set generated with the proposed affine matrix transformation technique in (14).

**M2**: Approximate set derived from hyper-box approaches [14], [15].

**M3**: Approximate set obtained with virtual-battery (VB) based approaches [18], [19].

**M4**: Approximate set produced by the traditional homothet-based polytope approaches [1], [6], [7]. Note that the standard approaches cannot handle coupled HVAC loads, but incorporating the proposed matrix transformation technique extends them to accommodate the approximation of $\mathbb{P}^{\mathrm{bld}}$.

The resulting approximations are shown in Fig. 6. Among the four, M2 is the most conservative. By neglecting inter-temporal coupling, it captures only 14.4% of the exact polytope's volume. M3 partially accounts for temporal flexibility, but due to the use of a predefined approximate energy storage model, it fails to capture the complex thermal coupling in multi-zone systems, representing only 39.4% of the true volume. In contrast, M4 incorporates multi-zone thermal parameters into the base set and uses homothetic transformations to partially capture thermal coupling, achieving 61.7% volume coverage. However, this geometric approach is limited in high-dimensional multi-period power space and cannot fully capture the complex thermal dynamics of multi-zone systems. In comparison, the proposed M1 applies a more general affine transformation to adapt the inner approximate polytope to the intricate coupling dynamics, recovering 89.5% of the exact volume. The above results highlight the accuracy of our approximation approach.

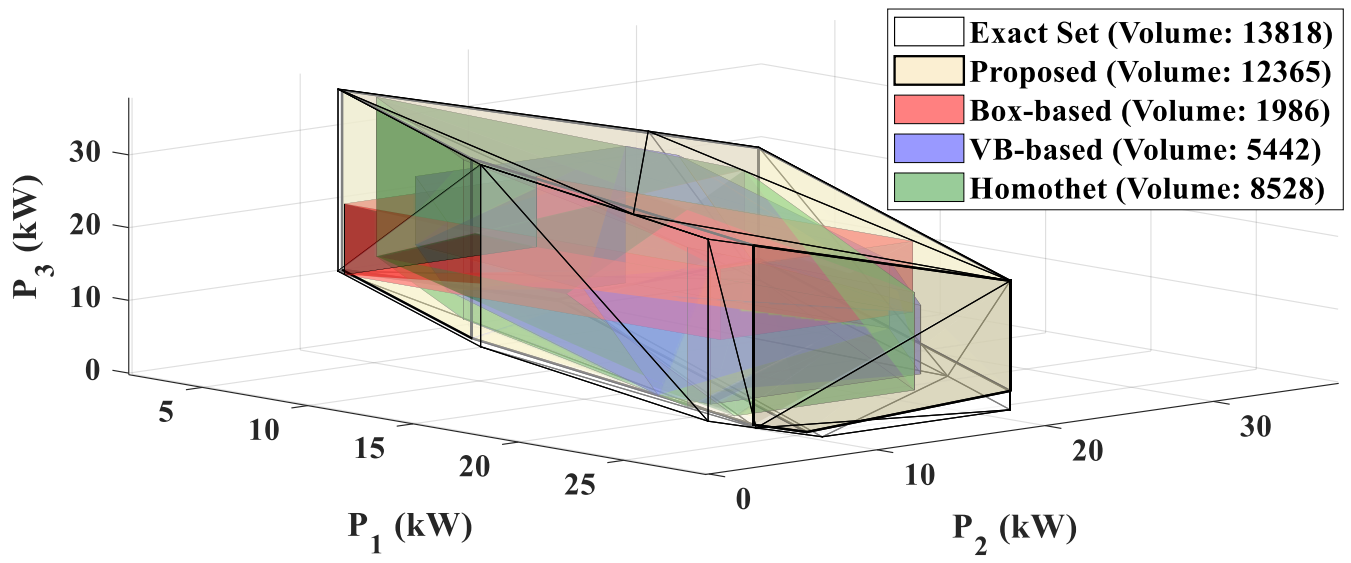

Fig. 6. Approximation results in the illustrative example.

*2) Aggregation Performance of Large-Scale Building HVAC Loads*: To validate the aggregator-level implementation, we integrate the aggregation approaches into the aggregator's day-ahead electricity cost minimization problem (25). As shown in TABLE II, 100 buildings participate in the aggregation, each containing 2 to 8 zones. We evaluate four cases, noting that the VB model is excluded from this comparison due to convergence failure within 24 hours when scaled to 100 buildings.

**Case 0**: Centralized control without aggregation. This serves as the ideal benchmark for evaluation.

**Case 1 (Proposed)**: Aggregation using the proposed two-level framework.

**Case 2**: Aggregation using the hyper-box approach [14], [15].

**Case 3**: Aggregation using traditional homothet-based polytope approach [1], [6], [7].

First, we present the 31-day scheduling results under varying electricity prices in Fig. 7. Case 1 achieves the lowest average UPR of 3.28% with the smallest fluctuations, consistent with our toy example findings. This confirms that our approach captures the most HVAC aggregate flexibility across different price scenarios. In contrast, Cases 2 and 3 show lower aggregation accuracy and larger UPR fluctuations. These approaches cannot fully capture complex thermal coupling, leading to conservative approximations that underestimate available flexibility.

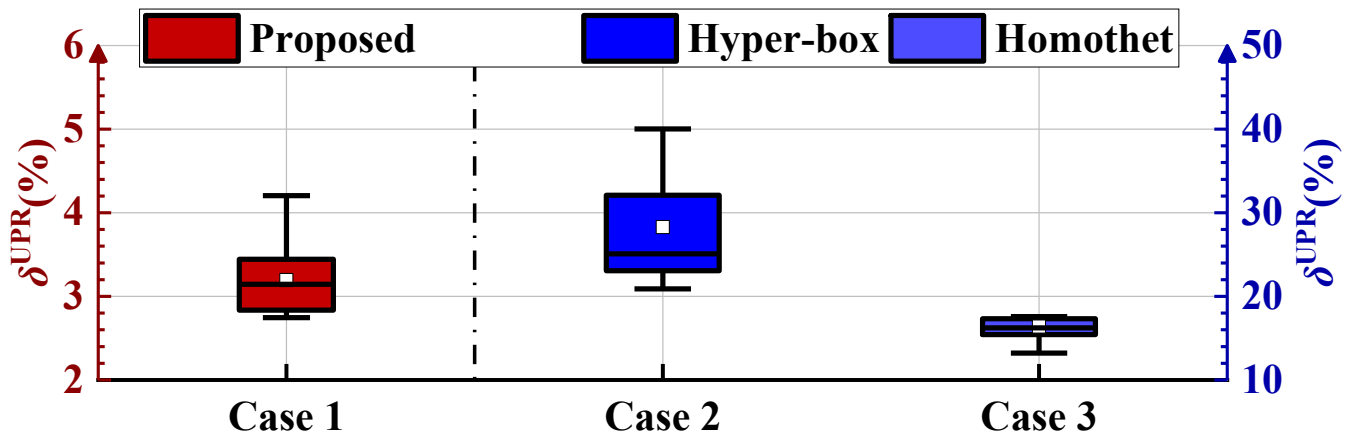

Fig. 7. Scheduling results in 31 days.

Next, we examine the July 9 scheduling results in detail. Fig. 8 shows the outdoor temperature and internal heat gain profiles for this day. Fig. 9 presents the aggregate power schedules. As expected, the aggregator purchases extra energy during low-price periods to charge thermal storage and discharges it during peak prices, minimizing daily costs. The Case 1 curve nearly overlaps with the benchmark, outperforming the other two cases and confirming that our aggregation preserves the true multi-zone thermal dynamics. Fig. 10 -Fig. 11 show the normalized power and temperature profiles in each zone using the proposed disaggregation strategy. All profiles remain within limits, confirming that users' thermal comfort constraints are satisfied and disaggregated power schedules are feasible.

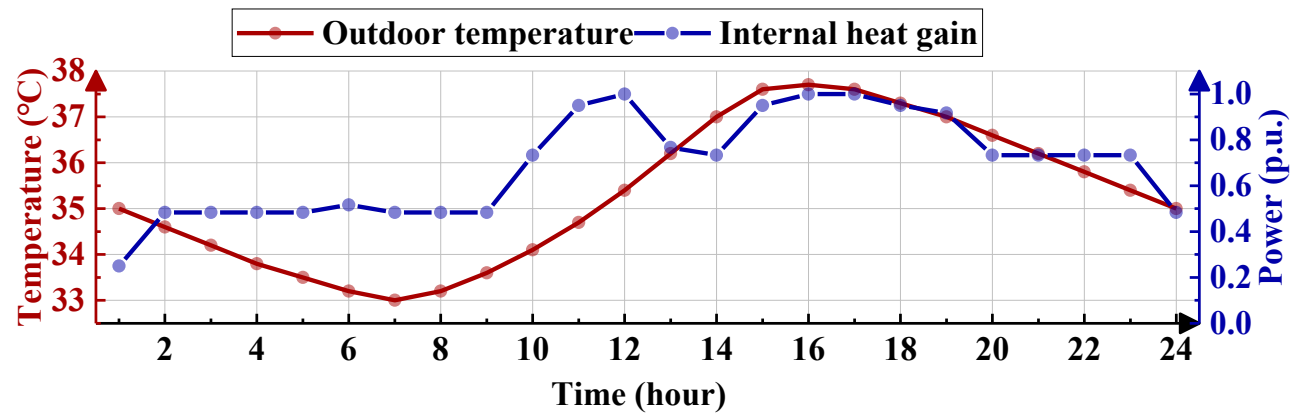

Fig. 8. Profiles of outdoor temperature and internal hear gain.

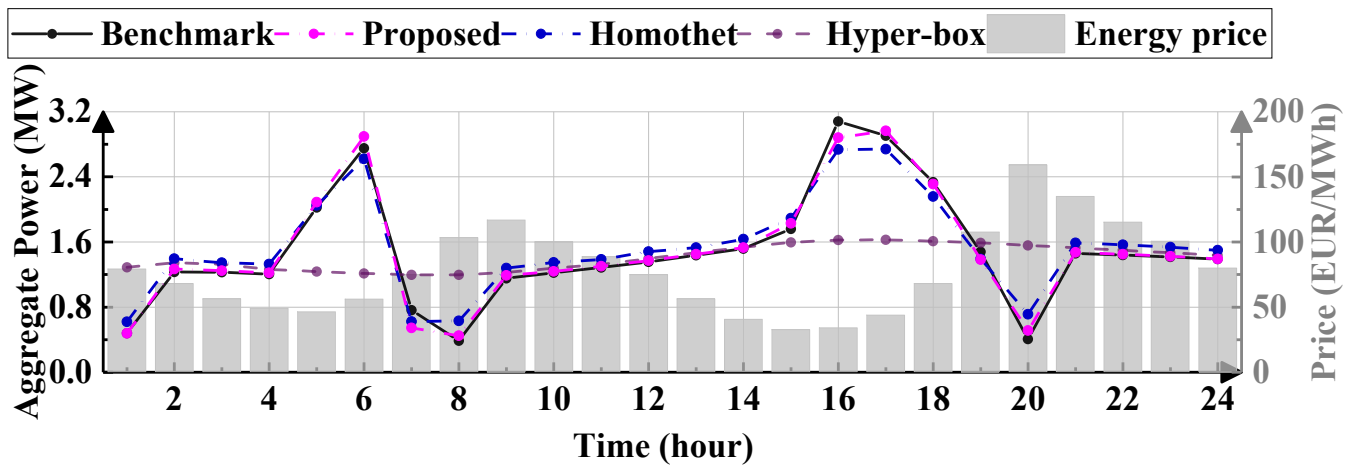

Fig. 9. Aggregate power profiles in one typical day.

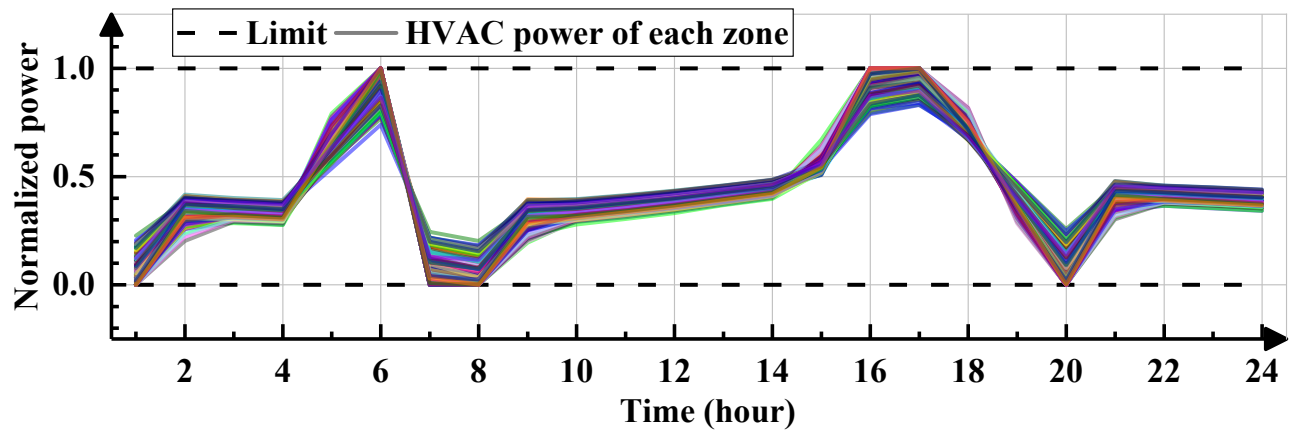

Fig. 10. Power profiles of all zones in Case 1 (*Normalized power* is defined as the $(P_{n,i,t}^{\text{hvac}} - P_{n,i}^{\text{hvac,min}})/(P_{n,i}^{\text{hvac,max}} - P_{n,i}^{\text{hvac,min}})$ which ranges from 0 to 1).

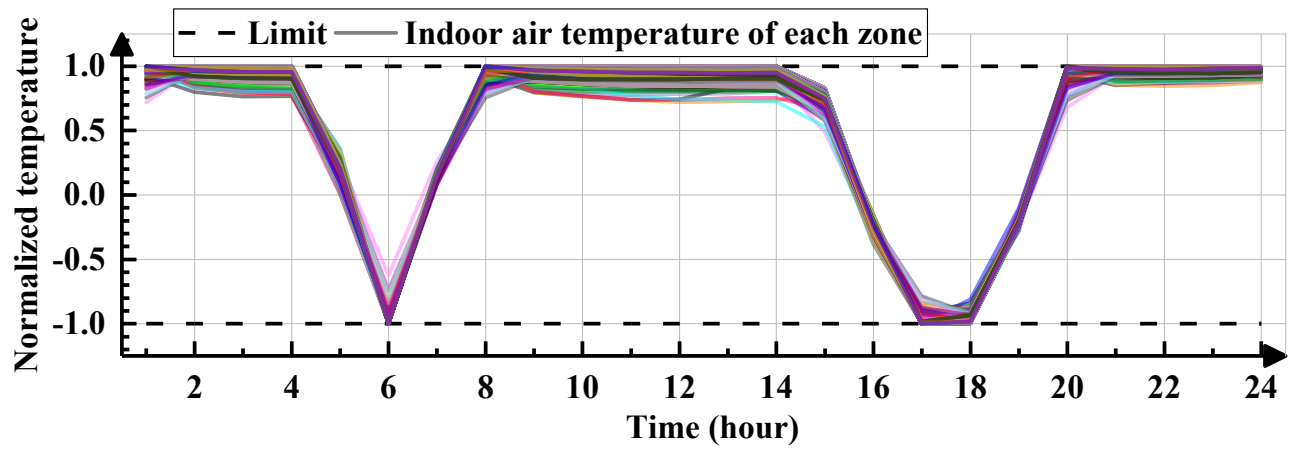

Fig. 11. Temperature profiles of all zones in Case 1 (*Normalized temperature* is defined as in (24.b) which ranges from -1 to 1).

Finally, we examine how aggregation performance varies with HVAC parameter distribution changes. Fig. 12 plots the empirical average UPR against the heterogeneity parameter $\alpha$ for each case. Even with small heterogeneity ($\alpha = 0.2$), the homothet approach performs poorly with an average UPR of 9.21%, indicating its inability to recover the true aggregate flexibility set. This poor performance occurs because complex thermal coupling between zones can significantly alter flexibility sets as the number of coupled zones varies, even with minimal thermal parameter heterogeneity. Traditional homothet-based approaches rely solely on scaling and translation—limited geometric operations that become increasingly conservative in high-dimensional spaces and fail to capture complex thermal dynamics. In contrast, our proposed approach employs generalized affine transformations that maintain strong geometric adaptability in high-dimensional spaces. As heterogeneity between individual flexibility sets increases, the proposed method consistently outperforms the other approaches, achieving an average UPR of approximately 5% when $\alpha = 1.4$. This demonstrates our method's ability to accurately recover true multi-zone-coupled flexibility sets under parameter variations.

Overall, these results validate that the proposed aggregation approach unlocks most HVAC aggregate flexibility to achieve economic scheduling results close to the ideal benchmark.

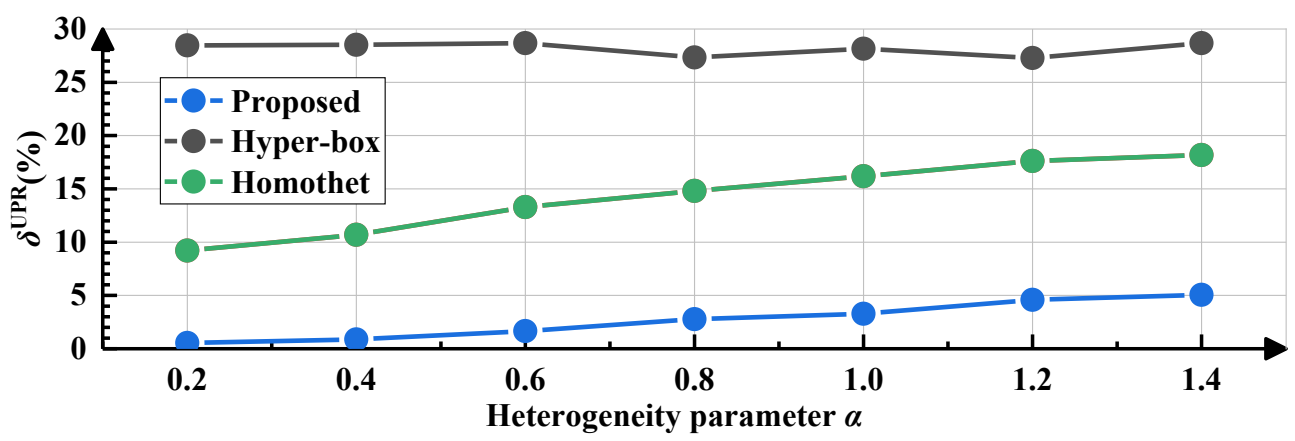

Fig. 12. Empirical average of the UPR $\delta^{\text{UPR}}$ versus the heterogeneity parameter $\alpha$ for three different inner approximation methods.

### C. *Effectiveness of Aggregation Reliability*

Next, we evaluate the reliability of our proposed aggregation approach under uncertainty using parameters from July 9.

*1) Impact of Uncertainty*: We first assess how forecast errors affect aggregate results. We generate an uncertainty dataset of 100 samples from a standard multivariate normal distribution with a diagonal covariance matrix, where each entry equals 5% of the corresponding forecast value (Fig. 8). Our data-driven $\varepsilon$-selection strategy yields $\varepsilon^* = 0.1171$. We compare two cases:

**Case 1a**: The proposed aggregation method without considering uncertainty.

**Case 1b**: The proposed aggregation method with the data-driven $\varepsilon$-selection strategy under uncertainty.

TABLE III presents the scheduling results. Case 1a, which ignores uncertainty, achieves higher response energy (14.248 MWh) by more aggressively utilizing HVAC flexibility. However, Case 1b, which incorporates uncertainty through our data-driven $\varepsilon$-selection strategy, intentionally reduces response energy by 12.8% (to 12.428 MWh) while increasing scheduling cost by 4.5%. This conservative approach deliberately reserves HVAC flexibility to hedge against forecast errors. Although Case 1a appears more economic in deterministic scenarios, its aggressive scheduling may become infeasible when forecast uncertainties materialize in practice.

To verify reliability, we conduct an out-of-sample test using 2,000 additional scenarios from the same distribution. Fig. 13-Fig. 14 show disaggregated indoor temperature profiles for a representative four-zone building, where solid red lines represent forecast trajectories and gray lines show realistic temperatures under uncertainty realizations. Fig. 13 reveals that the HVAC system in Case 1a will "hit the temperature boundary" under forecast conditions. When forecast errors occur, this scheduling leads to severe constraint violations, with a violation rate of 98.85% across the 2,000 test scenarios. This indicates that ignoring uncertainty results in overly aggressive aggregation. In contrast, our proposed approach maintains sufficient margin from temperature boundaries during scheduling (Fig. 14), achieving a violation rate of only 1.85% across the same test scenarios. These findings confirm that our approach produces reliable aggregate flexibility estimates and robust disaggregate schedules under uncertainty.

TABLE III
COMPARISONS OF UNCERTAINTY IMPACT

| Item | **Case 1a** | **Case 1b** |
|---|---|---|
| Scheduling cost (EUR) | 2303.7 | 2407.2 |
| Response energy* (MWh) | 14.248 | 12.428 |

**Response energy* is defined as the cumulative sum of absolute differences between the aggregator's scheduled power and baseline "no-control" power across all periods.

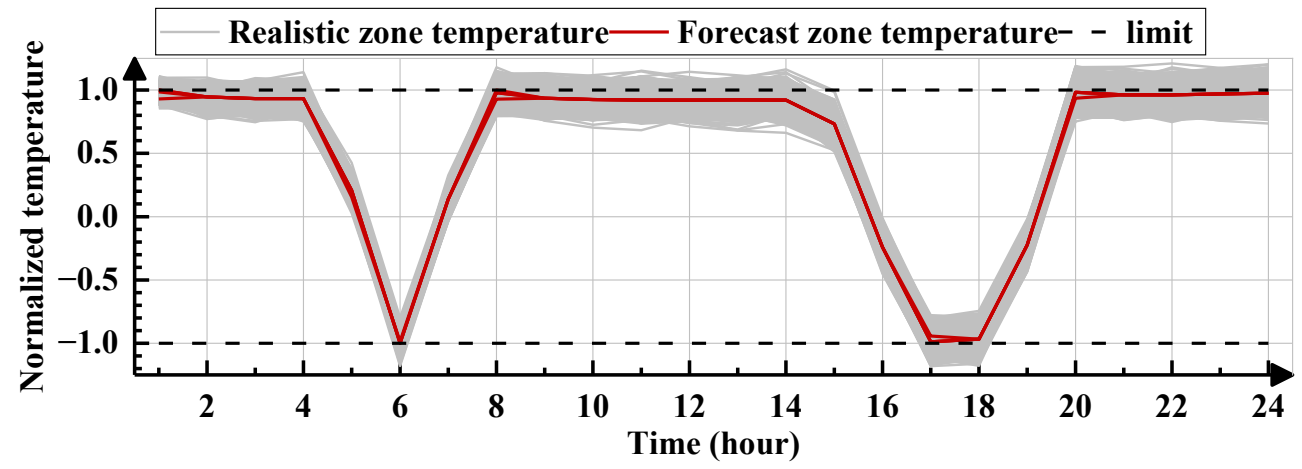

Fig. 13. Case 1a: Temperature profiles of four zones in a building under out-of-sample tests.

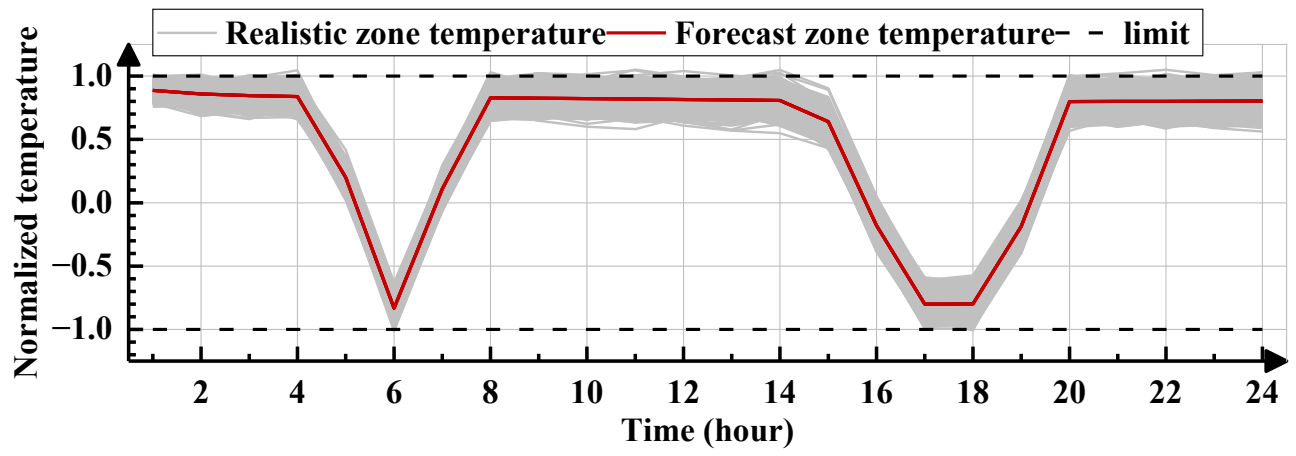

Fig. 14. Case 1b: Temperature profiles of four zones in a building under out-of-sample tests.

*2) Sensitivity Analysis of Risk Attitude*: Given forecast uncertainties, the aggregator's choice of allowable violation probability $\varepsilon$ significantly affects the aggregate result and scheduling performance. We examine how different $\varepsilon$ values impact Case 1b performance (Fig. 15). Results show that as $\varepsilon$ decreases (indicating greater risk aversion), response energy decreases while scheduling cost increases at an accelerating rate. This occurs because lower $\varepsilon$ values force the aggregator to report more conservative power and temperature bounds, which shrink the operational flexibility but provide greater protection against constraint violations. Conversely, higher $\varepsilon$ values allow more aggressive scheduling with greater response energy but increased violation risk. From Fig. 15, we observe that our data-driven $\varepsilon$-selection strategy $\varepsilon^* = 0.1171$ can identify a good balance that maximizes operational benefits while maintaining acceptable reliability levels for aggregators.

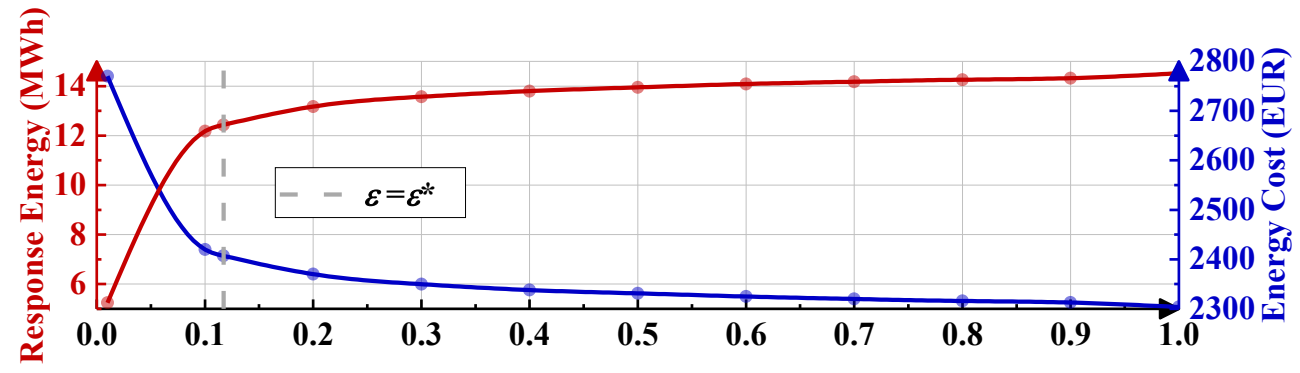

Fig. 15. Sensitivity analysis of risk attitude.

### D. Computational Efficiency Analysis

We conclude with an evaluation of computational efficiency. As noted in **Remark 2**, our approach can leverage parallel computing for large-scale aggregation, so overall runtime remains largely independent of the number of buildings. Instead, the inner-approximation step becomes the dominant cost, as its complexity grows with each building's zone count. As shown in TABLE IV, CPU time increases as the number of zones expands, but computational time remains under 4 minutes even for a 14-zone building, which is negligible compared to the 24-hour day-ahead scheduling horizon. This demonstrates the computational efficiency because the inner approximation problem (22.a)-(22.d) is a linear program and is readily solved by most optimization packages such as CPLEX or Gurobi. These solvers have been optimized to the point that a single linear program can be solved in the order of minutes even for large systems [38]. Therefore, our aggregation approach is applicable for practical large-scale implementations.

TABLE IV
INNER APPROXIMATION TIME WITH DIFFERENT ZONE SCALE

| Zone Number | Variable Number | Constraint Number | CPU Time (s) |
|---|---|---|---|
| 2 | 19,848 | 23,448 | 1.10 |
| 3 | 29,472 | 34,872 | 5.31 |
| 4 | 39,096 | 46,296 | 9.35 |
| 5 | 48,720 | 57,720 | 16.02 |
| 6 | 58,344 | 69,144 | 19.19 |
| 7 | 67,968 | 80,568 | 30.38 |
| 8 | 77,592 | 91,992 | 40.42 |
| 9 | 87,216 | 103,416 | 68.95 |
| 10 | 96,840 | 114,840 | 84.01 |
| 11 | 106,464 | 126,264 | 119.81 |
| 12 | 116,088 | 137,688 | 148.63 |
| 13 | 125,712 | 149,112 | 152.22 |
| 14 | 135,336 | 160,536 | 196.51 |

## V. CONCLUSION

This paper presents a two-level framework for coupling-aware aggregation of large-scale multi-zone HVAC loads under uncertainty. At the building level, we derive an analytical model that tracks uncertainty propagation across zones and provides the foundation for subsequent aggregator-level integration. At the aggregator level, we generalize existing polytope-based methods to perform inner approximations between polytopes of different dimensions, extending their applicability to complex multi-zone HVAC systems with thermal coupling. Simulation results demonstrate that our approach outperforms existing approaches by achieving high aggregation accuracy with low computational complexity. Also, our approach strikes an effective balance between robustness and conservatism, ensuring reliable flexibility estimates under uncertainty.

Future research will examine how additional real-world complexities affect HVAC load disaggregation, including heterogeneous response willingness across buildings and inter-building fairness considerations. Moreover, extending our framework to design more efficient algorithms for ultra-large-scale buildings with massive thermal zones presents an interesting research direction. Additionally, understanding the impact of individual HVAC parameter estimation errors on overall aggregation performance is an interesting topic for future investigation.

# VI. Appendix

## A. *Proof of* Proposition 1

The proof consists of three parts. First, we show that $\mathbb{P}_n^{\text{bld,mid}}(\boldsymbol{\xi}_n) \subseteq \mathbb{P}_n^{\text{bld,low}}(\boldsymbol{\xi}_n)$. Then, we demonstrate the reverse inclusion, $\mathbb{P}_n^{\text{bld,low}}(\boldsymbol{\xi}_n) \subseteq \mathbb{P}_n^{\text{bld,mid}}(\boldsymbol{\xi}_n)$. Finally, we establish the dimensionality relation clearly.

*1) Proof of* $\mathbb{P}_n^{\text{bld,mid}}(\boldsymbol{\xi}_n) \subseteq \mathbb{P}_n^{\text{bld,low}}(\boldsymbol{\xi}_n)$: Given the relationship $\mathbf{P}_n^{\text{bld}} = \boldsymbol{\Gamma}_n^{\text{ztb}}\mathbf{P}_n^{\text{hvac}}$, we have $[\mathbf{P}_n^{\text{hvac}}; \mathbf{P}_n^{\text{bld}}] = [\mathbf{P}_n^{\text{hvac}}; \boldsymbol{\Gamma}_n^{\text{ztb}}\mathbf{P}_n^{\text{hvac}}] = [\mathbf{I}_n^{\text{hvac}}; \boldsymbol{\Gamma}_n^{\text{ztb}}]\mathbf{P}_n^{\text{hvac}}$.

Since the matrix $[\mathbf{I}_n^{\text{hvac}}; \boldsymbol{\Gamma}_n^{\text{ztb}}]$ has full column rank, we can identify a maximal linearly independent subset of its rows. Because $\boldsymbol{\Gamma}_n^{\text{ztb}}$ itself has full row rank, we may select this subset to form an invertible square matrix containing $\boldsymbol{\Gamma}_n^{\text{ztb}}$, denoted by $\boldsymbol{\Gamma}_n^{\text{ct}} \in \mathbb{R}^{N_n^{\text{I}}N^{\text{T}} \times N_n^{\text{I}}N^{\text{T}}}$. Such a matrix $\boldsymbol{\Gamma}_n^{\text{ct}}$ can be explicitly determined via Gaussian-Jordan elimination. Consequently, the following holds: $\boldsymbol{\Gamma}_n^{\text{ct}}\mathbf{P}_n^{\text{bld}} = [\mathbf{P}_n^{\text{aux}}; \mathbf{P}_n^{\text{bld}}]$, where $\mathbf{P}_n^{\text{aux}} \in \mathbb{R}^{N_n^{\text{I}}N^{\text{T}} - N^{\text{T}}}$ represents the auxiliary variable vector. Substituting this into the inequality $\mathbf{A}_n^{\text{ieq3a}}\mathbf{P}_n^{\text{hvac}} + \mathbf{A}_n^{\text{ieq3b}}\boldsymbol{\xi}_n \le \mathbf{b}_n^{\text{ieq3}}$, we obtain $\mathbf{A}_n^{\text{ieq3a}}(\boldsymbol{\Gamma}_n^{\text{ct}})^{-1}[\mathbf{P}_n^{\text{aux}}; \mathbf{P}_n^{\text{bld}}] + \mathbf{A}_n^{\text{ieq3b}}\boldsymbol{\xi}_n \le \mathbf{b}_n^{\text{ieq3}}$.

This verifies that $\mathbb{P}_n^{\text{bld,mid}}(\boldsymbol{\xi}_n) \subseteq \mathbb{P}_n^{\text{bld,low}}(\boldsymbol{\xi}_n)$.

*2) Proof of* $\mathbb{P}_n^{\text{bld,low}}(\boldsymbol{\xi}_n) \subseteq \mathbb{P}_n^{\text{bld,mid}}(\boldsymbol{\xi}_n)$: According to $(\boldsymbol{\Gamma}_n^{\text{ct}})^{-1}[\mathbf{P}_n^{\text{aux}}; \mathbf{P}_n^{\text{bld}}]$, for any given vector $[\mathbf{P}_n^{\text{aux}}; \mathbf{P}_n^{\text{bld}}]$, there exists a corresponding $\mathbf{P}_n^{\text{bld}} \in \mathbb{R}^{N_n^{\text{I}}N^{\text{T}}}$ such that $\boldsymbol{\Gamma}_n^{\text{ct}}\mathbf{P}_n^{\text{bld}} = [\mathbf{P}_n^{\text{aux}}; \mathbf{P}_n^{\text{bld}}]$ holds. Since $\boldsymbol{\Gamma}_n^{\text{ct}}$ constitutes a basis for $\mathbb{R}^{N_n^{\text{I}}N^{\text{T}} \times N_n^{\text{I}}N^{\text{T}}}$ and includes $\boldsymbol{\Gamma}_n^{\text{ztb}}$, there must exist an auxiliary transformation matrix $\mathbf{M}_n = [[\mathbf{m}_n^{11}, \mathbf{m}_n^{12}]; [\mathbf{0}_n^{21}, \mathbf{I}_n^{22}]]$ satisfying $\mathbf{M}_n\boldsymbol{\Gamma}_n^{\text{ct}} = [\mathbf{I}_n^{\text{hvac}}; \boldsymbol{\Gamma}_n^{\text{ztb}}]$, where $\mathbf{m}_n^{11}$ and $\mathbf{m}_n^{12}$ are appropriately chosen matrix blocks. Multiplying both sides of $\boldsymbol{\Gamma}_n^{\text{ct}}\mathbf{P}_n^{\text{bld}} = [\mathbf{P}_n^{\text{aux}}; \mathbf{P}_n^{\text{bld}}]$ by $\mathbf{M}_n$ yields $\mathbf{M}_n[\mathbf{P}_n^{\text{aux}}; \mathbf{P}_n^{\text{bld}}] = [\mathbf{I}_n^{\text{hvac}}; \boldsymbol{\Gamma}_n^{\text{ztb}}]\mathbf{P}_n^{\text{bld}}$. Extracting the lower blocks from this equation, we obtain the relationship $\mathbf{P}_n^{\text{bld}} = \boldsymbol{\Gamma}_n^{\text{ztb}}\mathbf{P}_n^{\text{bld}}$. Substituting back into the inequality $\mathbf{A}_n^{\text{ieq3a}}(\boldsymbol{\Gamma}_n^{\text{ct}})^{-1}[\mathbf{P}_n^{\text{aux}}; \mathbf{P}_n^{\text{bld}}] + \mathbf{A}_n^{\text{ieq3b}}\boldsymbol{\xi}_n \le \mathbf{b}_n^{\text{ieq3}}$ yields $\mathbf{A}_n^{\text{ieq3a}}\mathbf{P}_n^{\text{hvac}} + \mathbf{A}_n^{\text{ieq3b}}\boldsymbol{\xi}_n \le \mathbf{b}_n^{\text{ieq3}}$.

This confirms $\mathbb{P}_n^{\text{bld,low}}(\boldsymbol{\xi}_n) \subseteq \mathbb{P}_n^{\text{bld,mid}}(\boldsymbol{\xi}_n)$.

*3) Dimensionality relation*: Initially, the dimension of the original variable is $\mathbf{P}_n^{\text{hvac}} \in \mathbb{R}^{N_n^{\text{I}}N^{\text{T}}}$. After the coordinate transformation, the auxiliary variable dimension reduces to $\mathbf{P}_n^{\text{aux}} \in \mathbb{R}^{N_n^{\text{I}}N^{\text{T}} - N^{\text{T}}}$. Thus, the dimensional difference satisfies $\dim(\mathbf{P}_n^{\text{hvac}}) - \dim(\mathbf{P}_n^{\text{aux}}) = N^{\text{T}} = \text{rank}(\boldsymbol{\Gamma}_n^{\text{ztb}})$.

The proof is thus complete. ■

## B. *Proof of* Proposition 2

**Lemma 1** (Linear encodings for polytope containment). Given two full-dimensional convex polytopes ( $\mathbb{P}^{\text{i}} = \{\mathbf{P}^{\text{i}} \in \mathbb{R}^{N^{\text{i}}} \mid \mathbf{H}^{\text{i}}\mathbf{P}^{\text{i}} \le \mathbf{h}^{\text{i}}\}$ and $\mathbb{P}^{\text{e}} = \{\mathbf{P}^{\text{e}} \in \mathbb{R}^{N^{\text{e}}} \mid \mathbf{H}^{\text{e}}\mathbf{P}^{\text{e}} \le \mathbf{h}^{\text{e}}\}$ ) and two affine transformations ($\boldsymbol{\Gamma}^{\text{i}}\mathbb{P}^{\text{i}} + \boldsymbol{\gamma}^{\text{i}}$ and $\boldsymbol{\Gamma}^{\text{e}}\mathbb{P}^{\text{e}} + \boldsymbol{\gamma}^{\text{e}}$), the inner containment relationship $\boldsymbol{\Gamma}^{\text{i}}\mathbb{P}^{\text{i}} + \boldsymbol{\gamma}^{\text{i}} \subseteq \boldsymbol{\Gamma}^{\text{e}}\mathbb{P}^{\text{e}} + \boldsymbol{\gamma}^{\text{e}}$ is ensured if there exist auxiliary variables $\mathbf{G} \in \mathbb{R}^{\dim(\mathbf{P}^{\text{e}}) \times \dim(\gamma^{\text{i}})}$, $\boldsymbol{\Lambda} \in \mathbb{R}_{+}^{\dim(\mathbf{h}^{\text{e}}) \times \dim(\mathbf{h}^{\text{i}})}$, and $\boldsymbol{\beta} \in \mathbb{R}^{\dim(\mathbf{P}^{\text{e}})}$ such that:

$$\boldsymbol{\Gamma}^{\text{i}} = \boldsymbol{\Gamma}^{\text{e}}\mathbf{G}, \boldsymbol{\gamma}^{\text{e}} - \boldsymbol{\gamma}^{\text{i}} = \boldsymbol{\Gamma}^{\text{e}}\boldsymbol{\beta} \tag{27.a}$$

$$\boldsymbol{\Lambda}\mathbf{H}^{\text{i}} = \mathbf{H}^{\text{e}}\mathbf{G} \tag{27.b}$$

$$\boldsymbol{\Lambda}\mathbf{h}^{\text{i}} \le \mathbf{h}^{\text{e}} + \mathbf{H}^{\text{e}}\boldsymbol{\beta} \tag{27.c}$$

**Lemma 1**, which a known result in the convex analysis of polytopes, constitutes an advanced extension of Farkas' lemma. Detailed proofs can be found in [34], [35]. From **Lemma 1**, (18.b)-(18.d) ensures $\boldsymbol{\gamma}_n^{\text{aff}} + \boldsymbol{\Gamma}_n^{\text{aff}}\mathbb{P}^{\text{base}} \subseteq \boldsymbol{\Gamma}_n^{\text{ltp}}\mathbb{U}_n^{\text{bld,low}}(\boldsymbol{\xi}_n)$ for each building-level full-dimensional polytopes $\mathbb{U}_n^{\text{bld,low}}(\boldsymbol{\xi}_n)$, thereby yielding $\boldsymbol{\Gamma}^{\text{agg}}\mathbb{P}^{\text{base}} + \boldsymbol{\gamma}^{\text{agg}} \subseteq \mathbb{P}^{\text{agg}}(\boldsymbol{\xi}_{\text{n}})$ according to (17) and (18.a).

The proof is thus complete. ■

## C. *Proof of* Proposition 3

Since $\varepsilon = \varepsilon^* = \frac{1}{1+\left(\max_{\boldsymbol{\xi}_n \in \mathbb{X}_n} \Upsilon(\boldsymbol{\xi}_n)\right)^2}$ is chosen, (28.a) holds:

$$\frac{\mathbf{A}_{n,r}^{\text{ieq4c}}(\boldsymbol{\xi}_n - \boldsymbol{\mu}_n)}{\sqrt{\mathbf{A}_{n,r}^{\text{ieq4c}}\boldsymbol{\sigma}_n\left(\mathbf{A}_{n,r}^{\text{ieq4c}}\right)^{\text{T}}}} \le \Upsilon(\boldsymbol{\xi}_n) \le \max_{\boldsymbol{\xi}_n \in \mathbb{X}_n} \Upsilon(\boldsymbol{\xi}_n) = \sqrt{\frac{1-\varepsilon^*}{\varepsilon^*}} \tag{28.a}$$

Multiplying through by $\sqrt{\mathbf{A}_{n,r}^{\text{ieq4c}}\boldsymbol{\sigma}_n\left(\mathbf{A}_{n,r}^{\text{ieq4c}}\right)^{\text{T}}}$ yields the key inequality (28.b) for $\forall \boldsymbol{\xi}_n \in \mathbb{X}_n$ and $\forall r$:

$$\mathbf{A}_{n,r}^{\text{ieq4c}}\boldsymbol{\xi}_n \le \mathbf{A}_{n,r}^{\text{ieq4c}}\boldsymbol{\mu}_n + \sqrt{\frac{1-\varepsilon^*}{\varepsilon^*}}\sqrt{\mathbf{A}_{n,r}^{\text{ieq4c}}\boldsymbol{\sigma}_n\left(\mathbf{A}_{n,r}^{\text{ieq4c}}\right)^{\text{T}}} \tag{28.b}$$

Combining (20) with the bound in (28.b) yields that (28.c) holds for $\forall \boldsymbol{\xi}_n \in \mathbb{X}_n$, which demonstrates that choosing $\varepsilon = \varepsilon^*$ offers abundant robustness against $\forall \boldsymbol{\xi}_n \in \mathbb{X}_n$.

$$\boldsymbol{\Lambda}_n^{\text{aux}}\mathbf{h}^{\text{base}} + \mathbf{A}_{n,r}^{\text{ieq4c}}\boldsymbol{\xi}_n \le \mathbf{b}_n^{\text{ieq4}} + [\mathbf{A}_n^{\text{ieq4a}}, \mathbf{A}_n^{\text{ieq4b}}]\boldsymbol{\beta}_n^{\text{aux}} \tag{28.c}$$

If one were to choose $\varepsilon > \varepsilon^*$, then $\sqrt{\frac{1-\varepsilon}{\varepsilon}} < \sqrt{\frac{1-\varepsilon^*}{\varepsilon^*}}$, and the inequality in (28.a) would fail for at least one $\boldsymbol{\xi}_n$, breaking feasibility. Conversely, taking $\varepsilon < \varepsilon^*$ yields $\sqrt{\frac{1-\varepsilon}{\varepsilon}} > \sqrt{\frac{1-\varepsilon^*}{\varepsilon^*}}$, so the constraints (28.c) become strictly over-satisfied, resulting in excessive conservatism.

The proof is thus complete. ■

## *D. Proof of* **Proposition 4**

First, recall the closed-form disaggregation formula $\mathbf{P}_n^{\text{bld},*} = \boldsymbol{\gamma}_n^{\text{aff}} + \boldsymbol{\Gamma}_n^{\text{aff}}(\boldsymbol{\Gamma}^{\text{agg}})^{-1}(\mathbf{P}^{\text{agg},*} - \boldsymbol{\gamma}^{\text{agg}})$, and the identities $\boldsymbol{\Gamma}^{agg} = \sum_n(\boldsymbol{\Gamma}_n^{\text{aff}})$, $\boldsymbol{\gamma}^{\text{agg}} = \sum_n(\boldsymbol{\gamma}_n^{\text{aff}})$. Summing over all buildings gives $\sum_n \mathbf{P}_n^{\text{bld},*} = \sum_n \boldsymbol{\gamma}_n^{\text{aff}} + (\sum_n \boldsymbol{\Gamma}_n^{\text{aff}})(\boldsymbol{\Gamma}^{\text{agg}})^{-1}(\mathbf{P}^{\text{agg},*} - \boldsymbol{\gamma}^{\text{agg}}) = \mathbf{P}^{\text{agg},*}$, so the aggregate power balance is satisfied.

Next, since $\mathbf{P}^{\text{agg},*} \in \mathbb{P}^{\text{app}} = \boldsymbol{\Gamma}^{\text{agg}}\mathbb{P}^{\text{base}} + \boldsymbol{\gamma}^{\text{agg}}$ yields that there exists $\mathbf{P}^{\text{base}} \in \mathbb{P}^{\text{base}}$ such that $\mathbf{P}^{\text{agg},*} = \sum_n(\boldsymbol{\Gamma}_n^{\text{aff}}\mathbf{P}^{\text{base}} + \boldsymbol{\gamma}_n^{\text{aff}})$. Thus we may equivalently choose $\mathbf{P}_n^{\text{bld},*} = \boldsymbol{\Gamma}_n^{\text{aff}}\mathbf{P}^{\text{base}} + \boldsymbol{\gamma}_n^{\text{aff}}$ which still satisfies $\sum_n \mathbf{P}_n^{\text{bld},*} = \mathbf{P}^{\text{agg},*}$. Finally, by construction $\boldsymbol{\gamma}_n^{\text{aff}} + \boldsymbol{\Gamma}_n^{\text{aff}}\mathbb{P}^{\text{base}} \subseteq \boldsymbol{\Gamma}_n^{\text{ltp}}\mathbb{U}_n^{\text{bld,low}}(\boldsymbol{\xi}_n)$ we have $\mathbf{P}_n^{\text{bld},*} \in \boldsymbol{\Gamma}_n^{\text{ltp}}\mathbb{U}_n^{\text{bld,low}}(\boldsymbol{\xi}_n)$ for each building.

The proof is thus complete. ■

## *E. Digestible Case under Dimensional Mismatch*

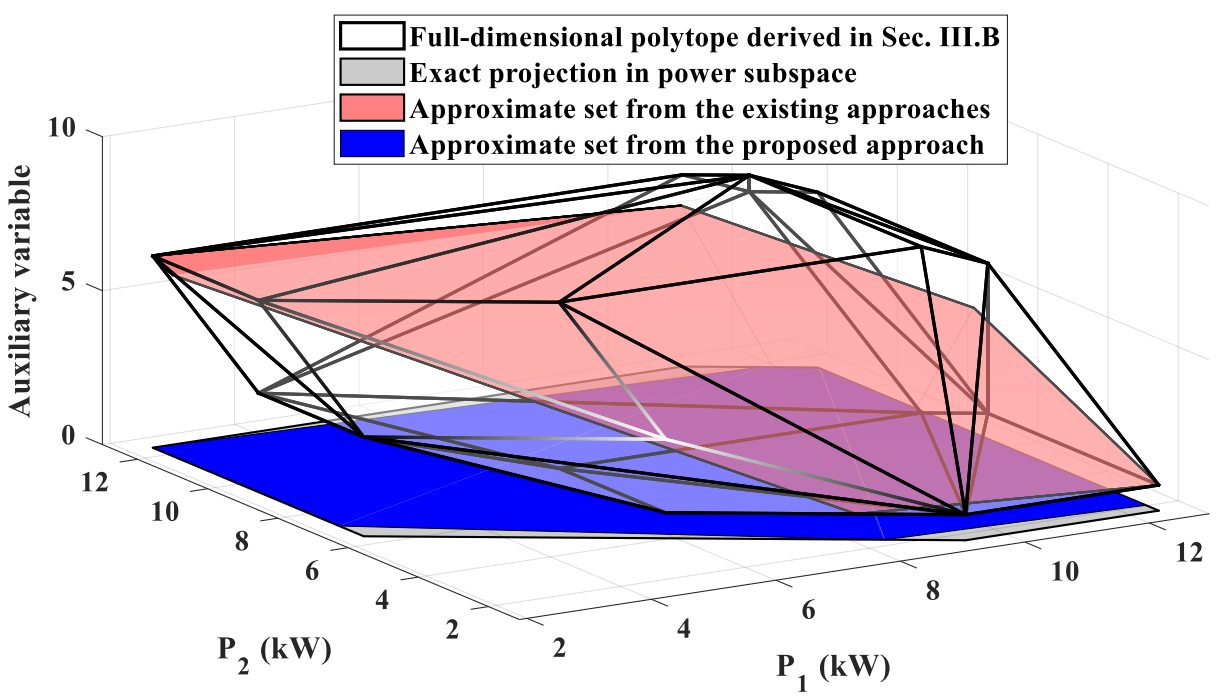

Fig. 16. Inner approximation under dimensional mismatch.

To demonstrate the advantages brought by the transition from fixed dimensions to differing dimensions, we provide a digestible case under dimensional mismatch. We consider a simple two-zone HVAC system over two scheduling periods and ignore uncertainty to highlight the main idea. The HVAC parameters are from TABLE II. Here, the full-dimensional polytope, derived in Section III.B as shown in (10), includes two aggregate power variables (denoted by $P_1$ and $P_2$) and two auxiliary variables (denoted by $P_1^{\text{aux}}$ and $P_2^{\text{aux}}$). For visualization, we eliminate $P_2^{\text{aux}}$ via the Fourier–Motzkin algorithm to produces the three-dimensional polytope. Our goal is to construct an inner approximation of the projection of this polytope onto the two-dimensional subspace spanned by $P_1$ and $P_2$. Fig. 16 compares results from existing fixed-dimension polytope approaches with those from the proposed differing-dimension approach. Both approaches guarantee the inner approximations. However, the fixed-dimension result is not guaranteed to stay strictly within the $P_1$-$P_2$ subspace, making it unsuitable for subsequent Minkowski-sum calculations in power space. Instead, the proposed matrix-transformation technique enforces this subspace constraint, ensuring a valid closed-form aggregation of multi-zone HVAC loads. This case study intuitively verifies the necessity and advantage of the proposed differing-dimension approach when aggregating high-dimensional, multi-zone-coupled HVAC flexibility.

## VII. Acknowledgment

We gratefully acknowledge Professor Cong Chen from Dartmouth College for her invaluable insights and discussions. We also thank the anonymous reviewers for their constructive feedback that greatly improved this paper.

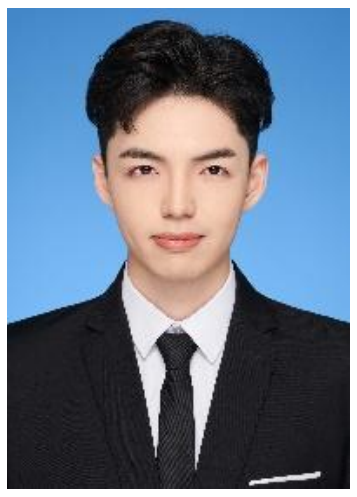

**Jingguan Liu** (Graduate Student Member, IEEE) received the bachelor's degree in electrical engineering from Huazhong University of Science and Technology, Wuhan, China, in 2022, where he is currently pursuing the Ph.D. degree, advised by Prof. Xiaomeng Ai and Prof. Jiakun Fang. Since September 2025, he has been a Visiting Student at the Dartmouth College, advised by Prof. Cong Chen. His research focuses on demand-side aggregation, energy storage operation, and optimization under uncertainty.

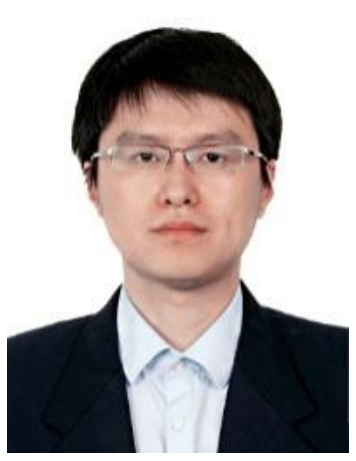

**Han Jiang** (Member, IEEE) received the B.S. and Ph.D. degrees from the College of Electrical Engineering, Zhejiang University, Hangzhou, China, in 2006 and 2012, respectively. He is currently working in Global Energy Interconnection Group Co., Ltd., Beijing. His research interests include power system planning, control and renewable energy.

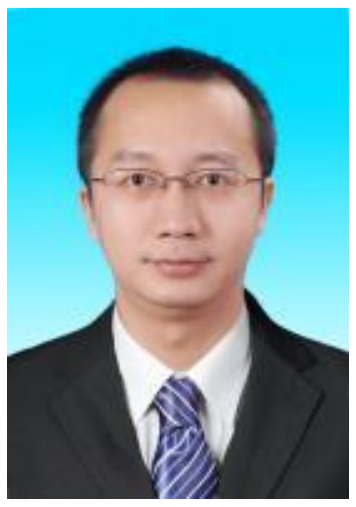

**Xiaomeng Ai** (Member, IEEE) received the B.Eng. degree in mathematics and applied mathematics and the Ph.D. degree in electrical engineering from the Huazhong University of Science and Technology (HUST), Wuhan, China, in 2008 and 2014 respectively. He is currently a Professor with the School of Electrical and Electronics Engineering, HUST. His research interests include robust optimization theory in power system, renewable energy integration, and integrated energy market.

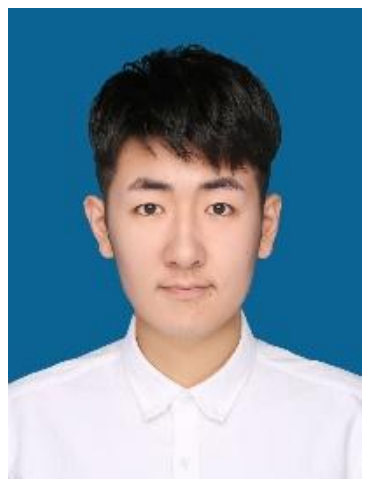

**Shengshi Wang** (Member, IEEE) received the B.Eng. degree in electrical engineering and automation from Chongqing University, Chongqing, China, in June 2020 and the Ph.D. degree in electrical engineering from Huazhong University of Science and Technology, Wuhan, China, in June 2025. From September 2023 to November 2024, he was a Visiting Student at Cardiff University, Cardiff, United Kingdom. Since September 2025, he has been a Research Fellow with the Singapore Institute of Technology, Singapore, which innovates closely with industry to pilot engineering solutions. His research interests lie in data-driven optimization—particularly robust optimization with decision-dependent uncertainties, approximate dynamic programming, and learning to optimize, with a focus on their applications in the energy systems.

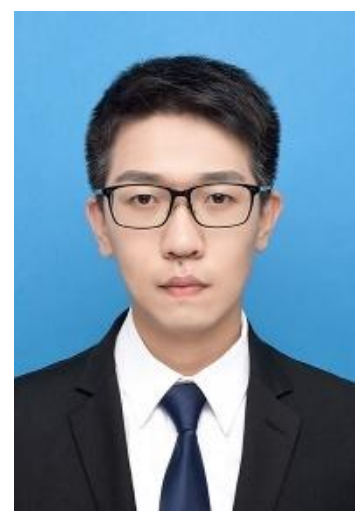

**Xizhen Xue** (Member, IEEE) received the B.E. and Ph.D. degrees in electrical engineering from the Huazhong University of Science and Technology, Wuhan, China, in 2019 and 2024, respectively. From 2023 to 2024, he was a Visiting Research Scholar with the Department of Electrical and Computer Engineering, Clarkson University, Potsdam, NY, USA. He is currently a Research Fellow with the School of Electrical and Electronic Engineering, Nanyang Technological University, Singapore. His current research interests include approximate dynamic programming, distributed optimization algorithm, energy storage scheduling and planning, and integrated energy systems.

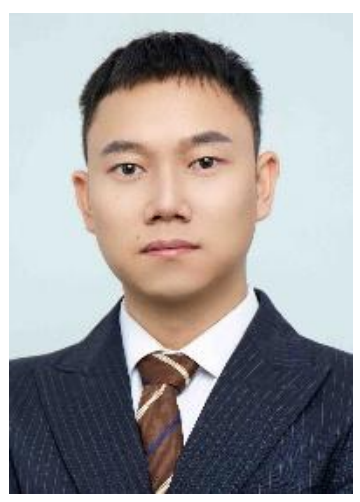

**Shichang Cui** (Member, IEEE) received the B.E. degree in automation and the Ph.D. degree in control science and engineering from the Huazhong University of Science and Technology (HUST), Wuhan, China, in 2016 and 2021, respectively. From 2019 to 2020, he was visiting the Department of Mechanical Engineering, University of Victoria, Canada, supported by the CSC Joint Doctoral Program. From 2021 to 2025, he was an assistant research fellow with the State Key Laboratory of Advanced Electromagnetic Technology, HUST, where he currently works as a Lecture. His current research interests include energy management for smart grids, distributed optimization, and game theory.

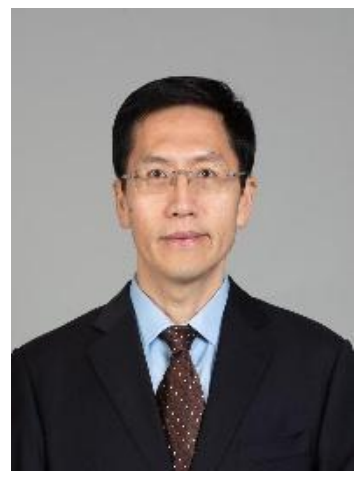

**Jinming Hou** received the Master degree in electrical engineering from the Norh China Electric Power University, Beijing, China, in 2006. He is currently a Senior Researcher with the Institute of Global Energy Interconnection Group Co., Ltd. He has been engaged in research work in the fields of power grid dispatching operation, multi-energy system planning, key technologies for energy transition, etc.

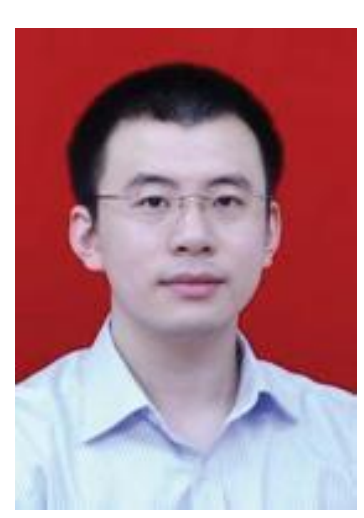

**Jiakun Fang** (Senior Member, IEEE) received the B.E. and Ph.D. degrees from the Huazhong University of Science and Technology (HUST), Wuhan, China, in 2007 and 2012, respectively. From 2012 to 2019, he was with the Department of Energy Technology, Aalborg University, Aalborg, Denmark. He is currently a Professor with the School of Electrical and Electronics Engineering, HUST. His research interests include the optimal integration of the power and gas systems, and the storage across multiple energy carriers.

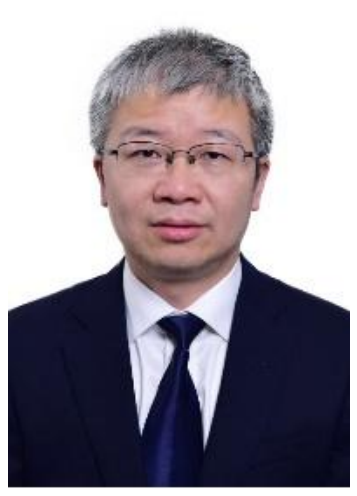

**Jinyu Wen** (Senior Member, IEEE) received the B.S. and Ph.D. degrees in electrical engineering from the Huazhong University of Science and Technology, Wuhan, China, in 1992 and 1998, respectively. He was a Visiting Student from 1996 to 1997, and a Research Fellow from 2002 to 2003, with the University of Liverpool, Liverpool, U.K., and a Senior Visiting Researcher with the University of Texas at Arlington, Arlington, TX, USA, in 2010. From 1998 to 2002, he was a Director Engineer with XJ Electric Company Ltd., China. In 2003, he joined HUST, where he is currently a Professor with the School of Electrical and Electronics Engineering. His current research interests include renewable energy integration, energy storage, multiterminal HVDC, and power system operation and control.